\documentclass[a4paper,11pt]{article}
\pdfoutput=1 
\usepackage{jinstpub}
\usepackage{graphicx} 
\usepackage{url}
\usepackage{lineno}

\newcommand{\Ig}  {I$_{\mathrm{G3U}}$}
\newcommand{\Ed}  {E$_{\mathrm{D}}$}
\newcommand{\Et}  {E$_{\mathrm{Transf}}$}
\newcommand{\Vg}  {V$_{\mathrm{GEM}}$}
\newcommand{\Dt}  {$D_{\mathrm{T}}$}
\newcommand{\Dl}  {$D_{\mathrm{L}}$}

\title{\boldmath Stability and detection performance of a GEM-based Optical Readout TPC with He/CF$_4$ gas mixtures}

\author[a,b]{E. Baracchini,} 
\author[c]{L. Benussi,}
\author[c]{S. Bianco,}
\author[c]{C. Capoccia,} 
\author[c,d]{M. Caponero,}
\author[e,f]{G. Cavoto,}
\author[a,b]{A. Cortez,}
\author[g]{I. A. Costa,}
\author[e]{E. Di Marco,}
\author[e]{G. D'Imperio,}
\author[a,b]{G. Dho,}
\author[e]{F. Iacoangeli,}
\author[c]{G. Maccarrone,}
\author[e,h]{M. Marafini,}
\author[c]{G. Mazzitelli,}
\author[e,f]{A. Messina,}
\author[g]{R. A. Nobrega,}
\author[c]{A. Orlandi,}
\author[c]{E. Paoletti,}
\author[c]{L. Passamonti,}
\author[i,j]{F. Petrucci,}
\author[c]{D. Piccolo,}
\author[c]{D. Pierluigi,}
\author[e,1]{D. Pinci,\note{Corresponding author.}}
\author[e]{F. Renga,}
\author[c]{F. Rosatelli,}
\author[c]{A. Russo,}
\author[c,k]{G. Saviano,}
\author[c]{and S. Tomassini}

\affiliation[a]{Gran~Sasso~Science~Institute,\\ L'Aquila, Italy}
\affiliation[b]{Istituto Nazionale di Fisica Nucleare,\\ Laboratori Nazionali del Gran Sasso, Assergi, Italy}
\affiliation[c]{Istituto Nazionale di Fisica Nucleare ,\\  Laboratori Nazionali di Frascati, Italy}
\affiliation[d]{ENEA Centro Ricerche Frascati, Frascati, Italy}
\affiliation[e]{Istituto~Nazionale~di~Fisica~Nucleare,\\ Sezione di Roma, Italy}
\affiliation[f]{Dipartimento di Fisica Sapienza Universit\`a di Roma, Italy} 
\affiliation[g]{Universidade Federal de Juiz de Fora, Juiz de Fora, Brazil}
\affiliation[h]{Museo Storico della Fisica e Centro Studi e Ricerche "Enrico Fermi",\\ Piazza del Viminale 1, Roma, Italy}
\affiliation[i]{Dipartimento di Matematica e Fisica, Universit\`a Roma TRE, Roma, Italy}
\affiliation[j]{Istituto Nazionale di Fisica Nucleare, Sezione di Roma TRE, Roma, Italy}
\affiliation[k]{Dipartimento di Ingegneria Chimica, Materiali e Ambiente, Sapienza Universit\`a di Roma, Roma, Italy}

\emailAdd{davide.pinci@roma1.infn.it}

\abstract{
The performance and long term stability of an optically readout Time Projection Chamber with an electron amplification structure based on three Gas Electron Multipliers was studied. He/CF$_4$ based gas mixtures were used in two different proportions (60/40 and 70/30) in a CYGNO prototype with 7 litres sensitive volume. 
With electrical configurations providing very similar electron gains, an almost full detection efficiency 
in the whole detector volume was found with both mixtures, while a light yield about 20\% larger for the 60/40 was found.
The electrostatic stability was tested by monitoring voltages and currents 
during 25 days. The detector worked in very stable and safe condition for the whole period. In the presence of less CF$_4$, a larger probability of unstable events was clearly detected.}

\keywords{Dark Matter detectors;
Gaseous imaging and tracking detectors;
Micropattern gaseous detectors;
Time projection Chambers.}

\begin{document}
\maketitle
\flushbottom

\section*{Introduction}
Liquid and gaseous Time Projection Chambers have been successfully proposed and exploited in last decades for very different applications, from High Energy Physics experiments on colliders to  the searches of Dark Matter (DM) massive particles.
In this latter field, a very promising technique involves the optical reading of gas electro-luminescence produced during the processes of electron multiplication. \cite{bib:ref1,bib:ref2,bib:ref3,bib:ref4,bib:nim_orange1,bib:jinst_orange2}. 

Thanks to the great progresses achieved in recent years in both the performance of Micro Pattern Gas Detectors and CMOS-based light sensors, optical readout provides several crucial advantages:
    
\begin{itemize}
\item the very good performance of optical sensors allows detection and reconstruction of very low energy releases;
\item sensors can be installed outside of the 
sensitive volume reducing the interference with the detector operation and possible sources of gas contamination;
\item the use of suitable lenses allows to acquire large surfaces with small sensors.
\end{itemize}

The CYGNO collaboration is developing the optical technique on Gas Electron Multipliers (GEM) \cite{bib:gem} 
working with He/CF$_4$ based gas mixtures, 
with the aim of realising a cubic meter demonstrator to study its performance for low mass Dark Matter directional search.
For such application, high detection efficiency and energy resolution are  needed together with a very good detector 
stability ensuring safe operation for long data takings.

In this paper, the performance of a CYGNO prototype 
is presented and discussed for the first time together with a detailed study of long term reliability  
in operating conditions with two different He/CF$_4$ mixtures (60/40 and 70/30).

\section{Experimental setup}

\subsection{LEMON detector}
\label{sect:lemon}
\begin{figure}[ht]
\centering
\includegraphics[width=0.9\textwidth]{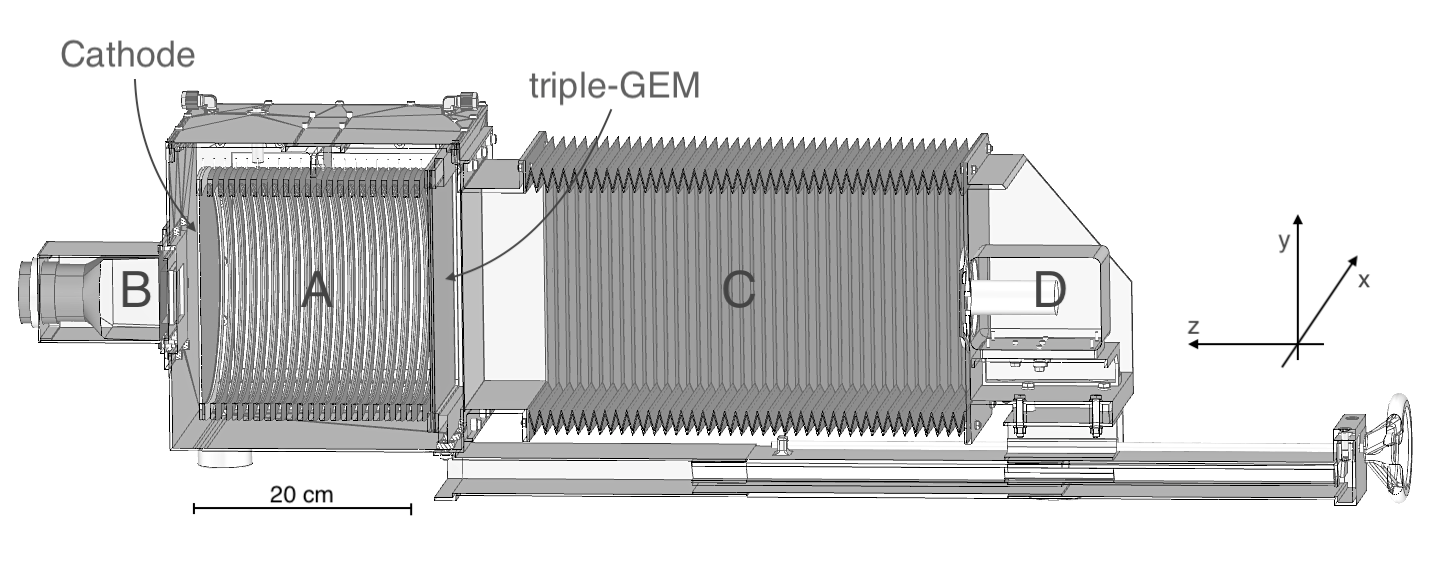}
\caption{Drawing of the experimental setup. In particular, the elliptical field cage closed on one side by the triple-GEM structure and on the other side by the semitransparent cathode (A), the PMT (B), the adaptable bellow (C) and the CMOS camera with its lens (D).} \label{fig:lemon}
\end{figure}

All studies presented in this paper have been carried out with the {\it LEMON} prototype~\cite{bib:eps, bib:ieee17, bib:ieee18}.
A sketch of this detector (described in more details in \cite{bib:fe55}) is shown in Fig~\ref{fig:lemon}.
It has a 7 litres sensitive volume (A) with a $20~cm$ long drift gap surrounded by an elliptical field cage closed on one side by a semitransparent cathode and on the other side by a~$20~\times~24~cm^2$ triple GEM structure.
Light produced in multiplication channels, is acquired by:
\begin{itemize}
    \item a photo-multiplier\footnote{Photonics XP3392} with 5~ns rise-time, a maximum QE of 12\% for 420~nm and a 76~$\times~76$~mm$^2$ square-window (trough the cathode);
    \item a scientific-CMOS based camera\footnote{Hamammatsu ORCA Flash 4}) with 2048~$\times$~2048 pixels with an active area of 6.5~$\times$~6.5~$\mu$m$^2$ each, equipped with a Schneider lens with 25~mm focal length and 0.95 aperture at a distance of 52.5~cm. The sCMOS sensor provides a quantum efficiency of about 70\% at 600~nm.
\end{itemize}
 
The drift volume was filled with He/CF$_4$ based gas mixtures. 
The operating performance with the two different gas mixtures (60/40 and 70/30) was studied.
According to previous studies, 
electro-luminesce spectra of He/CF$_4$ based mixtures show two main peaks: one around 300~nm and one around 620~nm. The relative light production depends on the relative amount of the two components in the gas mixture: the height of second peak increases with respect to the first one for a larger CF$_4$ percentage in the mixture~\cite{bib:Fraga}.

\subsection{Operating conditions}
\label{sec:oper}
The typical working configuration of the detector is: 
\begin{itemize}
    \item a gas flux of 200 cc/min;
    \item an electric field within the sensitive volume \Ed~=~0.5~kV/cm;
    \item an electric field in the 2~mm wide gaps between the GEMs \Et~= 2.5~kV/cm;
    \item a voltage difference across each GEM \Vg~=~460~V while operating with 60/40 mixture and \Vg~=~425~V while operating with 70/30 mixture.
\end{itemize}

According to results presented in \cite{bib:roby}, the behavior of electron gain with
the two gas mixtures as a function of \Vg\ can be described as:

\begin{eqnarray}
    {\mathrm{G}}_{60/40} (V_{\mathrm{Gem}}) = 28.36 \times e^{0.0235 \cdot V_{\mathrm{Gem}}\mathrm{(V)}} \\
    {\mathrm{G}}_{70/30} (V_{\mathrm{Gem}}) = 22.82 \times e^{0.0258 \cdot V_{\mathrm{Gem}}\mathrm{(V)}}
\end{eqnarray}

Therefore, since:
\begin{equation}
{\mathrm{G}}_{60/40}(460~\mathrm{V})/{\mathrm{G}}_{70/30}(425~\mathrm{V}) = 1.065
\end{equation}

the two values chosen for \Vg\ should allow to operate with very similar charge gains.

After two days of conditioning, the detector was closed in a 5 cm thick lead box, in order to partially shield it from soft cosmic rays and external natural radioactivity background. A slit on the box side, allowed to irradiate the sensitive volume with 5.9~keV photons produced by a 740~MBq $^{55}$Fe source at a $y$ position corresponding to half height of the field cage and free to move along $z$ axis.

\subsection{Data Acquisition}
\label{sec:daq}

In the events presented in this paper the sCMOS sensor 
was operated in auto-trigger mode with an exposure time of 40~ms.
The PMT waveform was sent into a 4 GS/s digitizer.
Only in the case that during the CMOS exposure window PMT waveform exhibited an amplitude larger than 80~mV it was acquired for a time interval of 25~$\mu$s and the image was stored.

In all tested configurations, {\it runs} with 1000 events were acquired.

In order to monitor and study the working conditions of the detector, voltage outputs and currents drawn by all high voltage channels supplying the different electrodes of the detectors, were logged with a sampling rate of 1 Hz. The used high voltage supply module is able to measure only negative currents. For this reason, it was possible to acquire only the currents provided to GEM and field cage cathodes.

\section{Operation stability}

\begin{figure}[ht]
	\centering
	\includegraphics[width=0.79\linewidth]{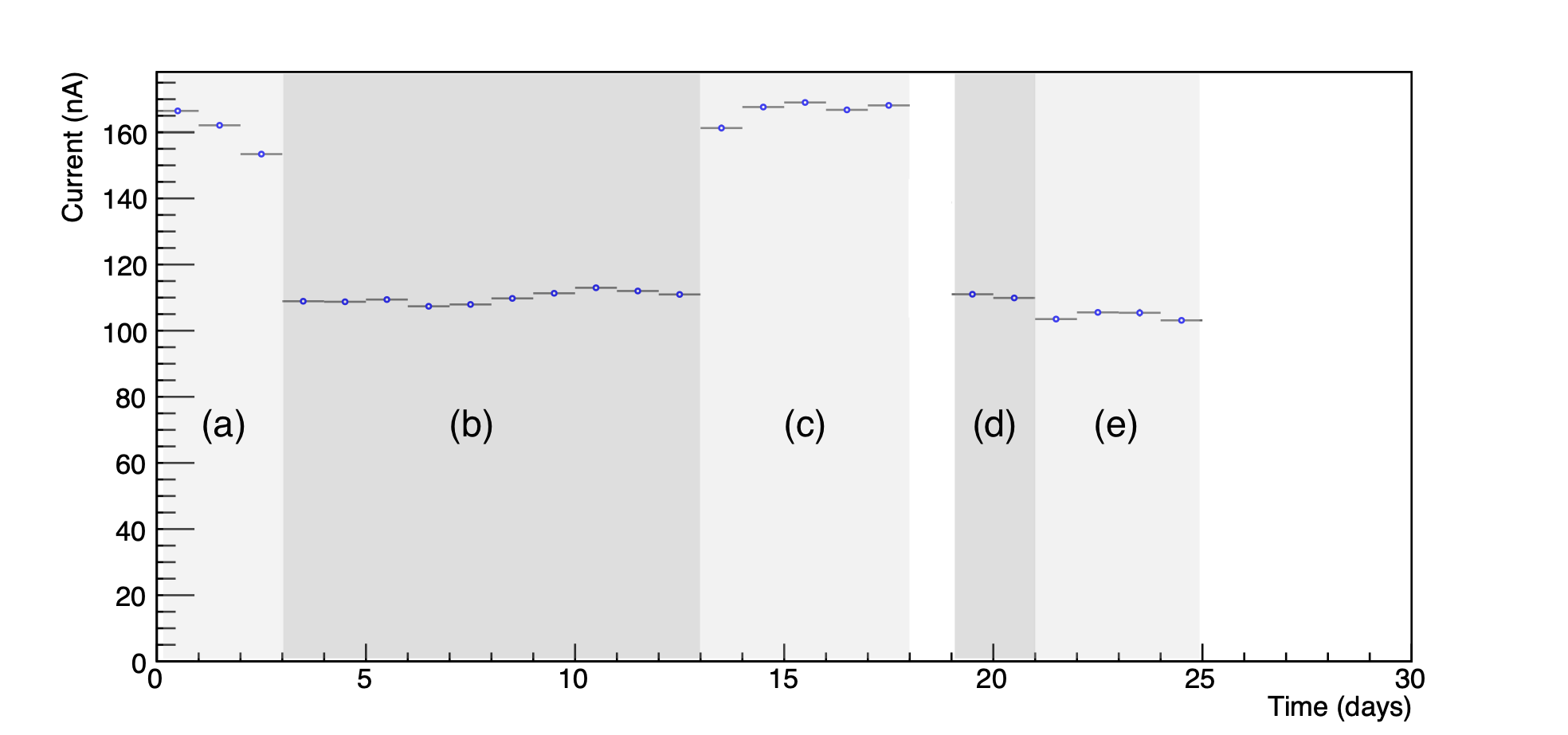} 
  	\caption{Average current drawn by the upper electrode of the third GEM \Ig\ as a function of time. Details about the experimental set-up in different periods are summarised in Table~\ref{tab:period}.}
  	\label{fig:period}
\end{figure}

Detector operational stability was evaluated during a 25 days long test. 
Figure~\ref{fig:period} shows the behavior of current drawn by the high voltage channel supplying the upper electrode (the one which faces the second GEM) of the third GEM in the stack (\Ig) averaged day by day. On this electrode, a fraction of ions produced during the avalanche is collected and thus \Ig\ is proportional to the charge produced by ionization in the sensitive volume amplified by the triple-GEM gain.  
For \Ed\~=~0, so that with no ionization charge collected on the GEM, it was found to be compatible with the sensitivity of current measurement (20~nA).

The data were collected for five consecutive periods, whose main characteristics are summarised in Tab~\ref{tab:period}
\begin{table}[h!]
\caption{Description of main settings in different periods of test}
\begin{center}
\begin{tabular}{ |r|c|c|c|c|c| } 
 \hline
 Period & Gas Proportion & Pb        & $^{55}$Fe & Collimator & Avg. Current\\ 
        & (He/CF$_4$)    & Shielding & Source    &            & (nA)\\ 
 \hline
 \hline
 {\bf (a)} & 60/40 & No & No & No  & 164 $\pm$ 2\\
 {\bf (b)} & 60/40 & Yes & No & No & 110 $\pm$ 1\\
 {\bf (c)} & 60/40 & Yes & Yes & No & 168 $\pm$ 2 \\
 {\bf (d)} & 60/40 & Yes & Yes & Yes & 110 $\pm$ 1\\
 {\bf (e)} & 70/30 & Yes & Yes & Yes & 104 $\pm$ 2\\
 \hline
\end{tabular}
\end{center}
\label{tab:period}
\end{table}

The 5~cm thick lead shield reduced the level of ionization due to natural background by about 30\% ({\bf (a)} and {\bf (b)} in Fig.~\ref{fig:period}). 
The ratio between the current drawn in periods {\bf (d)} and {\bf (e)} results to be 1.06~$\pm$~0.02 in good agreement with the value 1.065 expected for these operating conditions (see Sect.~\ref{sect:lemon}).

During normal operation, even if for most of the time the currents drawn by the detector electrodes were very stable, 
from time to time spikes occurred indicating detector instabilities.

Two different kinds of electrostatic instabilities have been observed:

\begin{itemize}
\item {\bf Hot spots.} Appearance of small luminous spots on the GEM surface as the one shown in Fig.\ref{fig:hotspot}, usually accompanied by negligible current increase.
\begin{figure}[ht]
	\centering
	\includegraphics[width=0.29\linewidth]{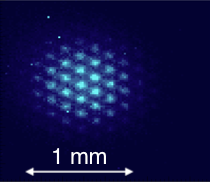}
	\includegraphics[width=0.60\linewidth]{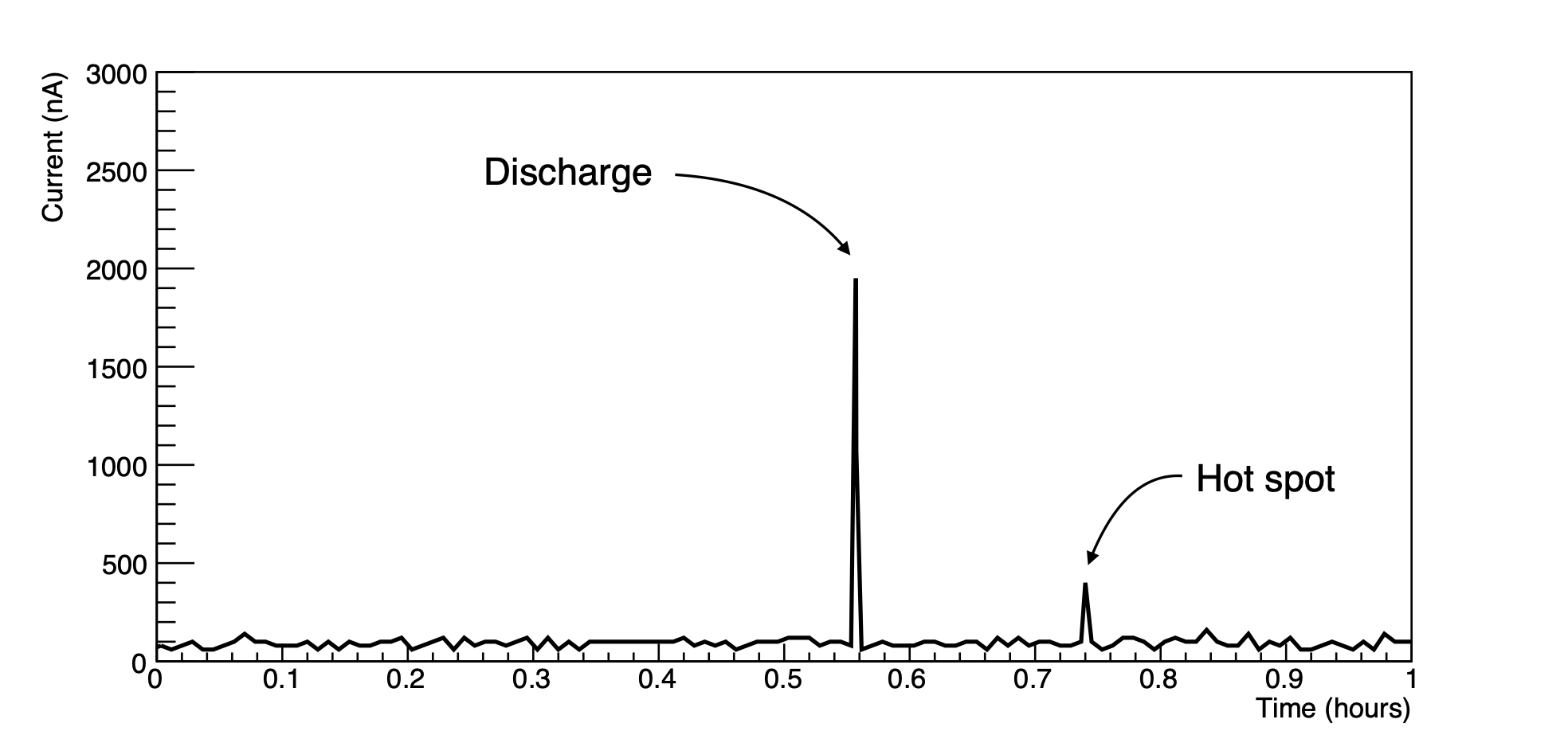}\\
\hspace{-2.5cm} \mbox{a)} \hspace{6.5cm} \mbox{b)}
\vspace{-0.3cm}

  	\caption{Zoomed image of a luminous spot on the GEM surface ({\it hot spot}) very likely due to self-sustaining micro-discharge occurring in one or more GEM channels (a). The structure of GEM channels is clearly visible. Example of the behavior of current drawn by the upper electrode of the third GEM \Ig\ as a function of time (b). The occurrence of a {\it discharge} and a {\it hot spot} are indicated (see text for details).}
  	\label{fig:hotspot}
\end{figure}

Some of these spots fades out with the time but in some case they start to slowly grow up (on a time scale of minutes). At some point they could even involve a current drawn by the GEM as large as to be measurable by the power supply (tens of nA, as shown on the right of Fig.~\ref{fig:hotspot}). These are probably due to self-sustaining micro-discharges happening in one or few GEM channels. It was found that a decrease of the voltage across all GEMs of 100~V is enough to dump this self-sustaining process and that, when \Vg\ is restored the hot-spot do not re-appears. An automatic hot-spot dump procedure was implemented to decrease all \Vg\ if the drawn current increases above some threshold. The \Vg\ is thus restored in 5 steps of 20 V with a 30 sec pause between each step. This procedure lasts about 3~minutes and introduce a {\it dead time} in the detector operation.

\item{\bf Discharges} High charge density due to very high ionizing particles or charge accumulation on electrode imperfections can suddenly discharge across GEM channels. In these events, a sudden increase in the drawn current is recorded with a voltage restoring on the electrodes through protection resistors on a few seconds time basis (an example is shown on the left of Fig.~\ref{fig:hotspot}). Also these events trigger the recovery procedure.
Even if these events are less frequent than hot spots 
they can be dangerous for the GEM structure and the energy released in the discharge can, in principle, damage it.

\end{itemize}

To study the occurrence of above events for the two gas mixtures, the detector currents were acquired while all other setting parameters were kept in stable and operational mode. 
For the 40/60 mixture the LEMON behavior has been acquired for 130 hours (5.4 days). In the whole period a total of 66 hot spots and 31 discharges were recorded giving an occurrence probability of 12.2 hot-spots/day and 5.7 discharges/day.
While operated with the 70/30 gas mixture, the detector showed a less stable conditions and therefore a shorted test time was needed to reach statistically significant results. In 60 hours of test (2.5 days), 121 hot-spots and 39 discharges occurred, corresponding to a probability of 48.4 hot-spots/day and 15.6 discharges/day.

\begin{figure}[ht]
\begin{center}
    
	\includegraphics[width=0.45\linewidth]{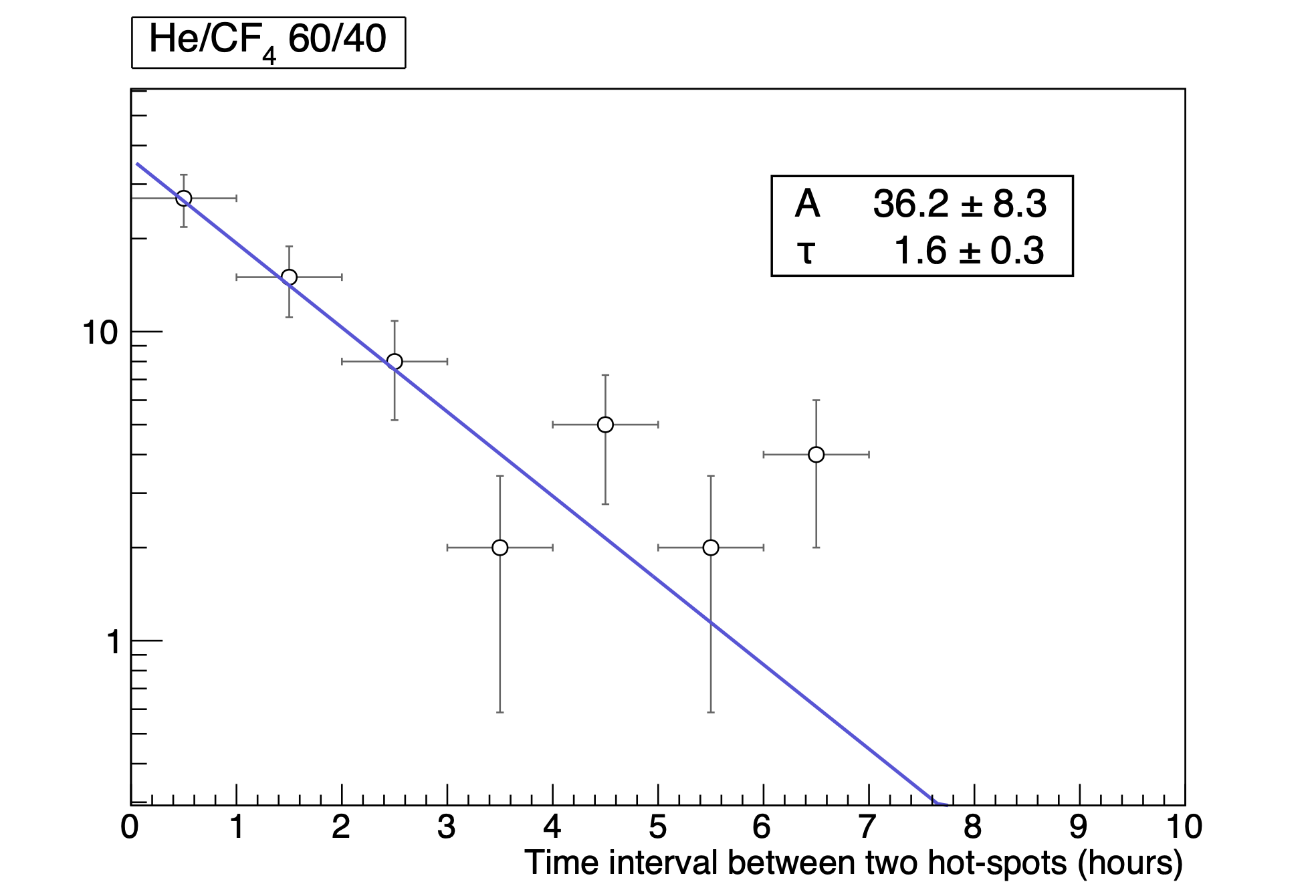}
	\includegraphics[width=0.45\linewidth]{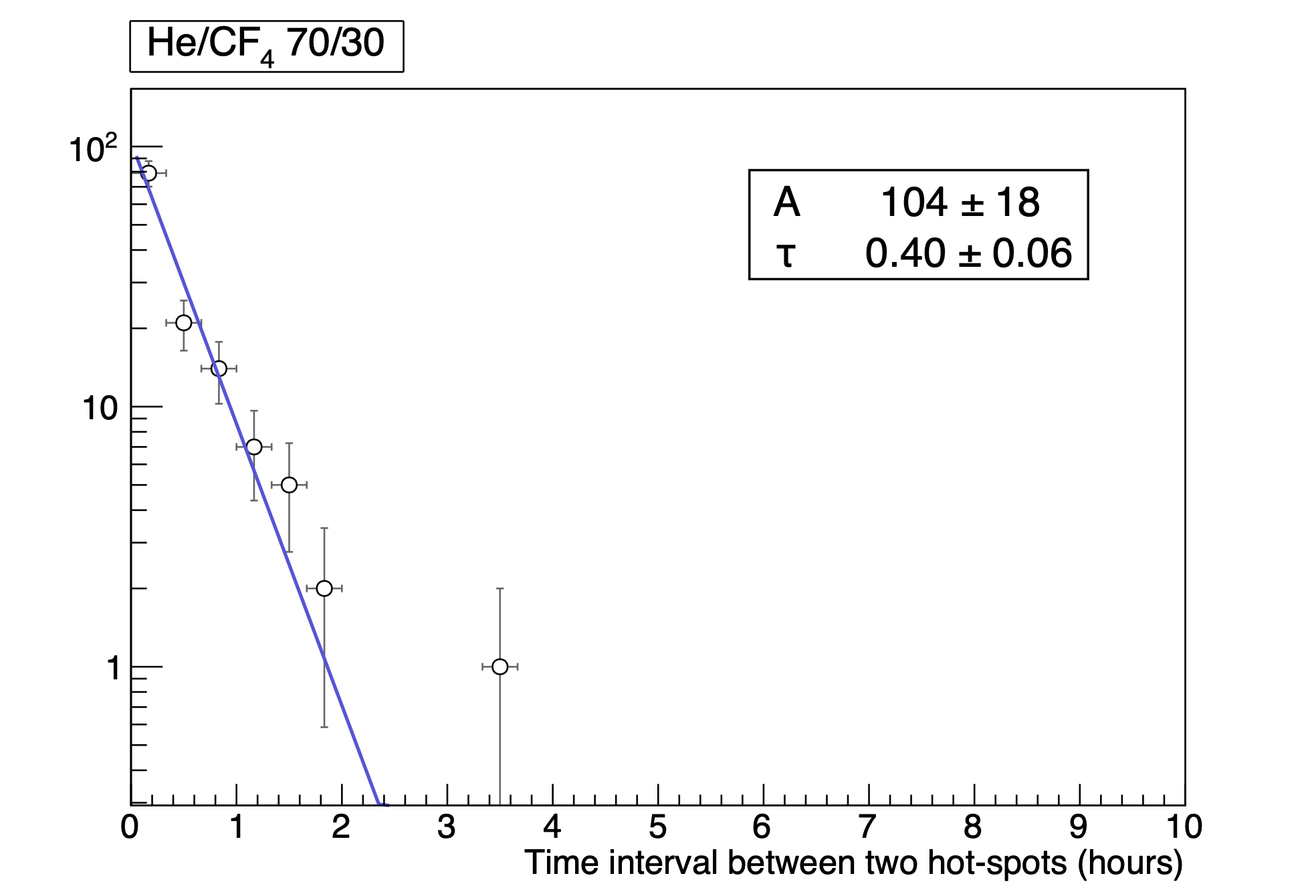}\\

\end{center}	
\vspace{-0.6cm}
\hspace{4cm}\mbox{a)} \hspace{6.5cm} \mbox{b)}
\vspace{-0.3cm}

  	\caption{Distributions of the interval between the appearance of two subsequent hot-spots (a) for He/CF$_4$ (60/40) and (b) for He/CF$_4$ (70/30)) with superimposed fits to exponential decrease.}
  	\label{fig:hDist}
\end{figure}

Figure~\ref{fig:hDist} shows the distributions of the time interval between the appearance of two subsequent hot-spots for the two gas mixtures.

The superimposed exponential fit to a function:
\begin{equation}
\label{eq:exp}
y = A~e^{-\frac{\Delta t}{\tau}}
\end{equation}

demonstrates that this distribution is the one expected for events occurring at completely random moments without any evident correlation~\cite{bib:knoll}.
In particular, the results of the fit confirm that the rate of hot-spots in He/CF$_4$ (70/30) mixture is 4 times larger than He/CF$_4$ (60/40).

\begin{figure}[ht]
	\begin{center}
	\includegraphics[width=0.45\linewidth]{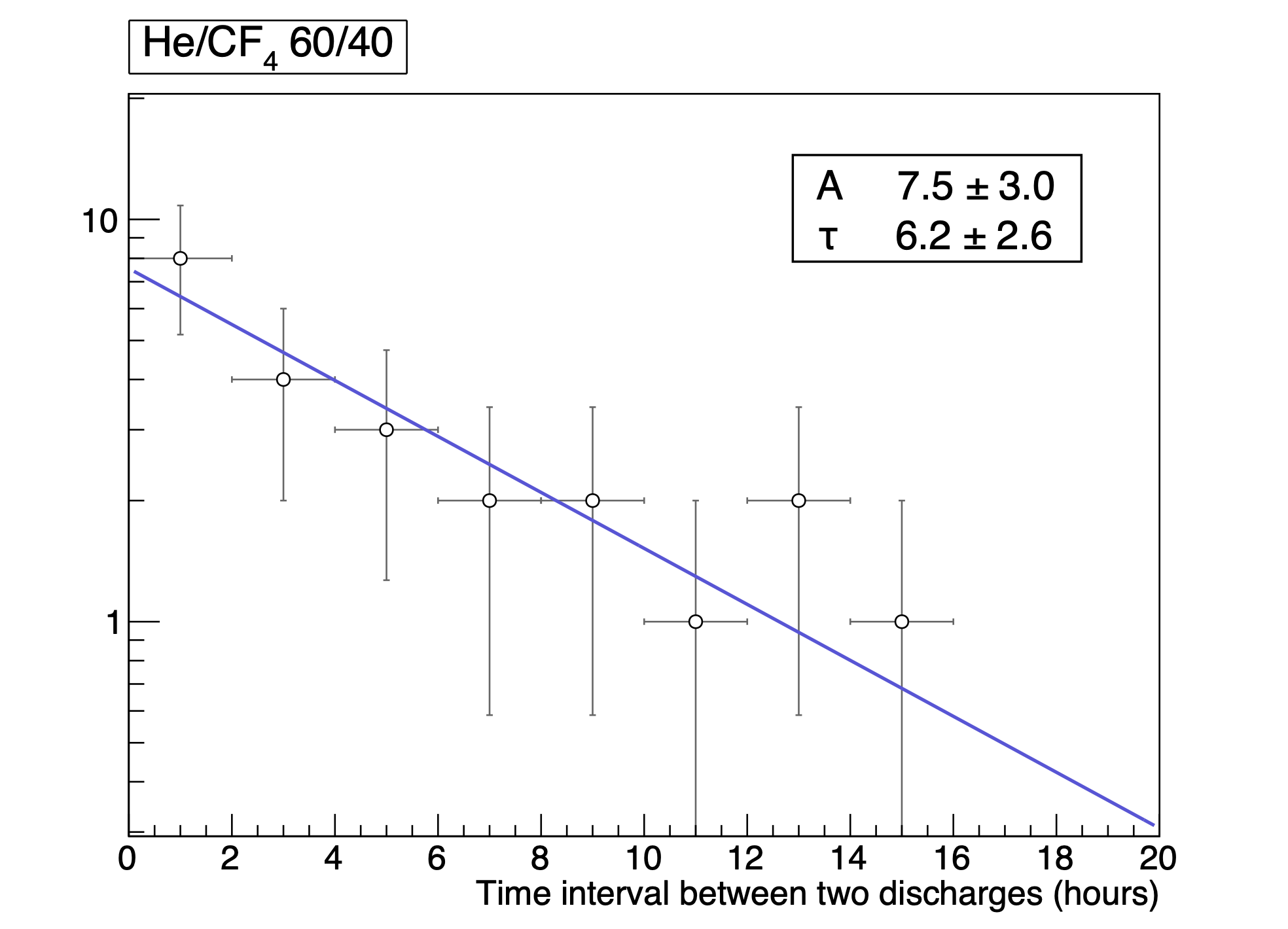}
	\includegraphics[width=0.45\linewidth]{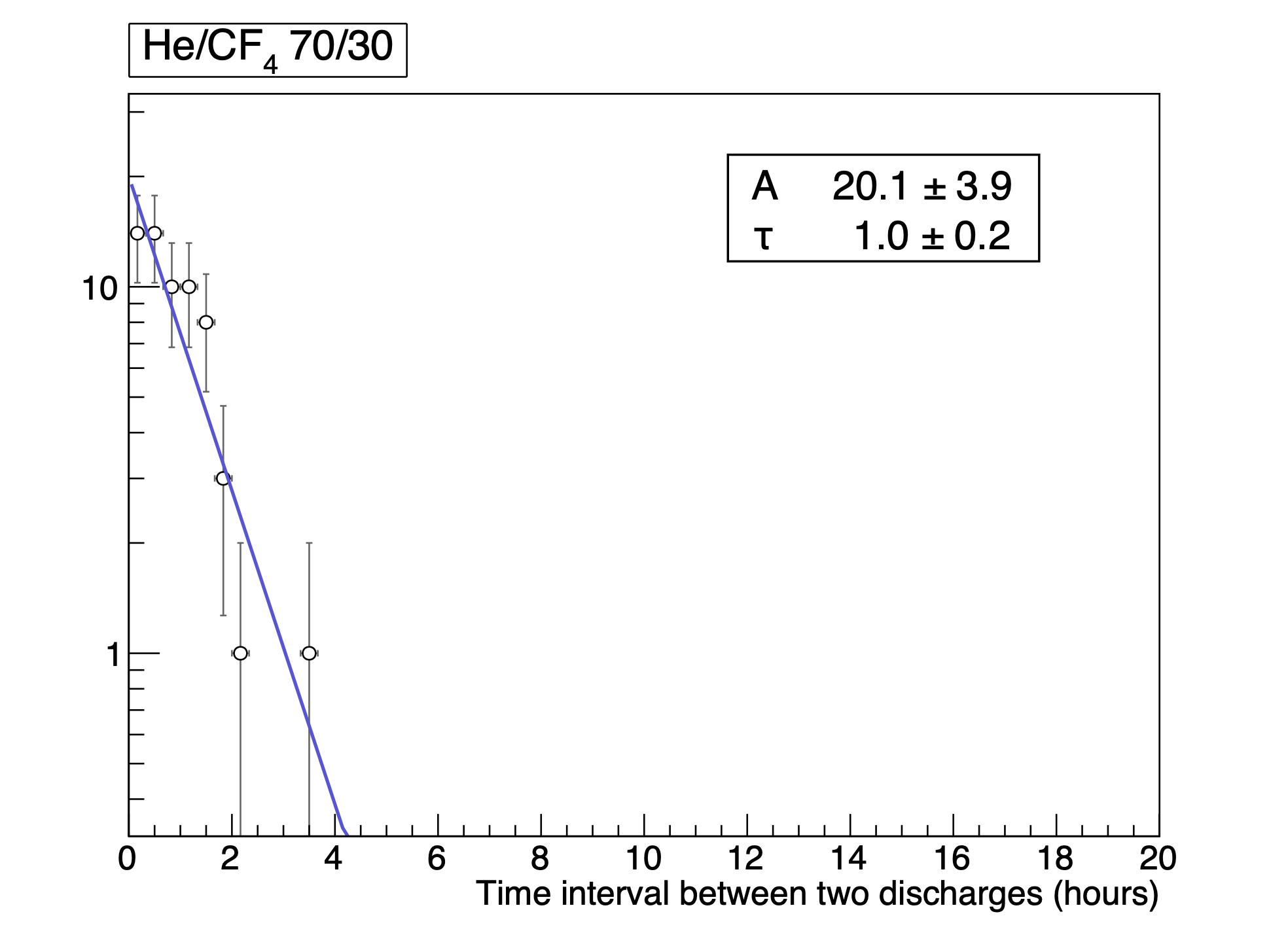}
	\end{center}	
\vspace{-0.6cm}
\hspace{4cm}\mbox{a)} \hspace{6.5cm} \mbox{b)}
\vspace{-0.3cm}

  	\caption{Distributions of the interval between the occurrence of two subsequent discharges (a) for He/CF$_4$ (60/40) and (b) for He/CF$_4$ (70/30)) with superimposed fits to exponential decrease.}
  	\label{fig:hDistDis}
\end{figure}

As it was done for the hot-spots, the distributions of the time intervals between the occurrence of two subsequent discharges was studied for both the mixtures (Fig.~\ref{fig:hDistDis}).

The behavior is well described by the function in \ref{eq:exp} showing that also the occurrence of these events is random in time without any evident correlation.
The results of the fit confirm that the probability of having a discharge in He/CF$_4$ (70/30) mixture is 6 times larger than He/CF$_4$ (60/40).

It results evident that a lower amount of CF$_4$, able to quench and keep under control possible production of large amount of charge during the multiplication processes, resulted in a less stable electrostatic configuration. The instability events gave rise to a detection inefficiency due to dead time introduced by recovering procedures of 3.8\% (60/40) and 13.3\% (70/30).

Nevertheless, it is important to outline that, in both cases, the detector behavior seemed not to be critical and the provided performance was satisfactory. 

\section{Simulation of gas mixtures}
\label{sect:simu}
The parameters of the two gas mixtures relevant to study the electron transport in the field cage were calculated
by means of Garfield \cite{bib:garfield1,bib:garfield2}.

Because electrons diffuses along their drift in the field cage, when they arrive on the GEM,
they trigger avalanche multiplications of an area larger than the ionization region. 
The photons emitted by gas electro-luminescence in the avalanches will thus create the light-spots on the sensor (see for example on the right of Fig.~\ref{fig:spot}).

After a drift over a distance $z$, transverse and longitudinal profiles of electron clouds produced in these spots can be described by Gaussian curves with standard deviations that can be calculated as:

\begin{eqnarray}
\label{eq:diff}
\sigma_{\mathrm{T}} = \sqrt{\sigma^2_{\mathrm{T0}} \oplus D^2_{\mathrm{T}} \cdot z} \\
\sigma_{\mathrm{L}} = \sqrt{\sigma^2_{\mathrm{L0}} \oplus D^2_{\mathrm{L}} \cdot z}
\end{eqnarray}

where $\sigma_{T0}$ and $\sigma_{L0}$ are constant contributions due diffusion in the GEM structure and to channel pitch and \Dt\ and \Dl\ are {\it transverse and longitudinal diffusion coefficients} that depend on the gas mixture and the electric field.

\begin{figure}[ht]
\centering
\includegraphics[width=0.4\textwidth]{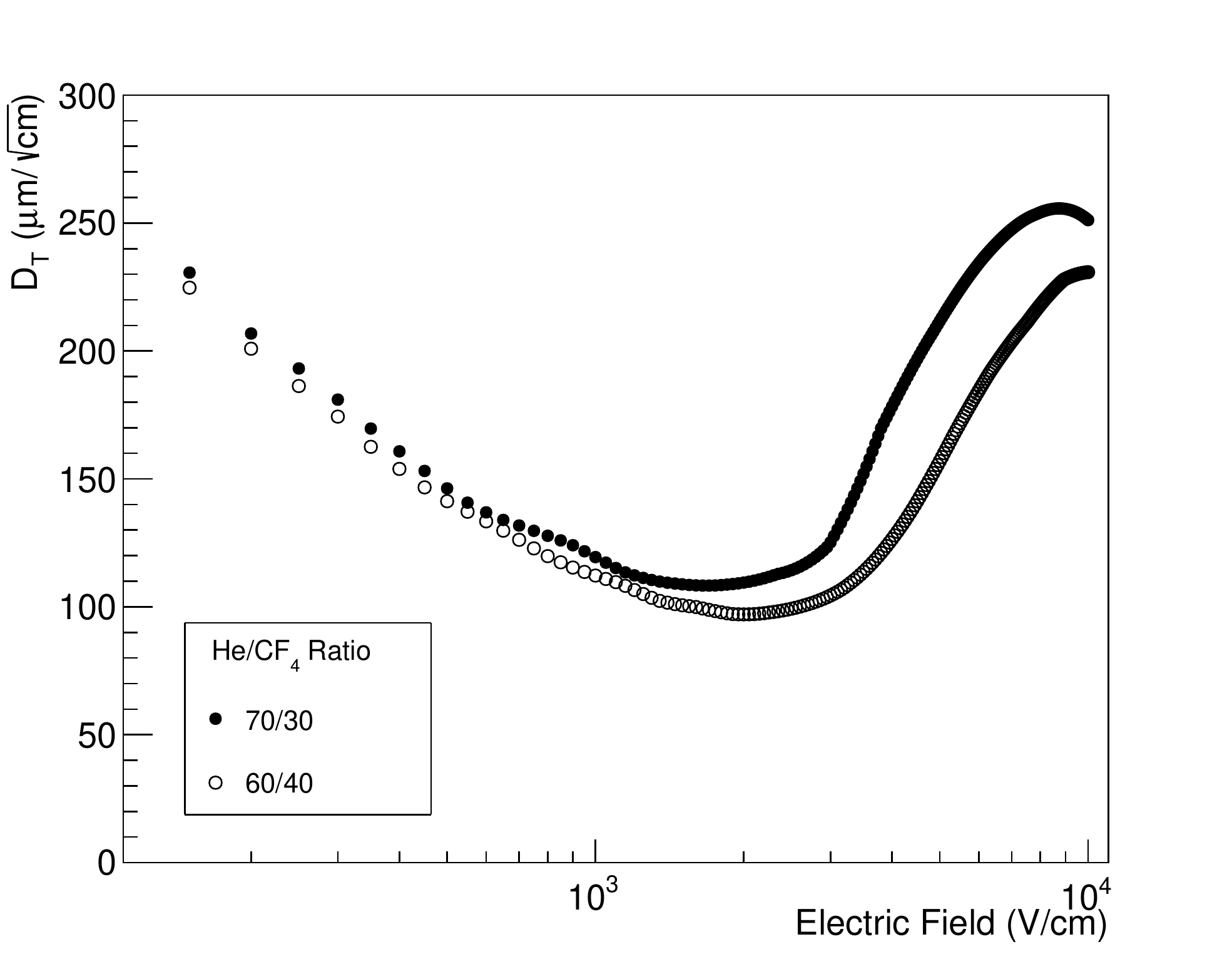}
\includegraphics[width=0.4\textwidth]{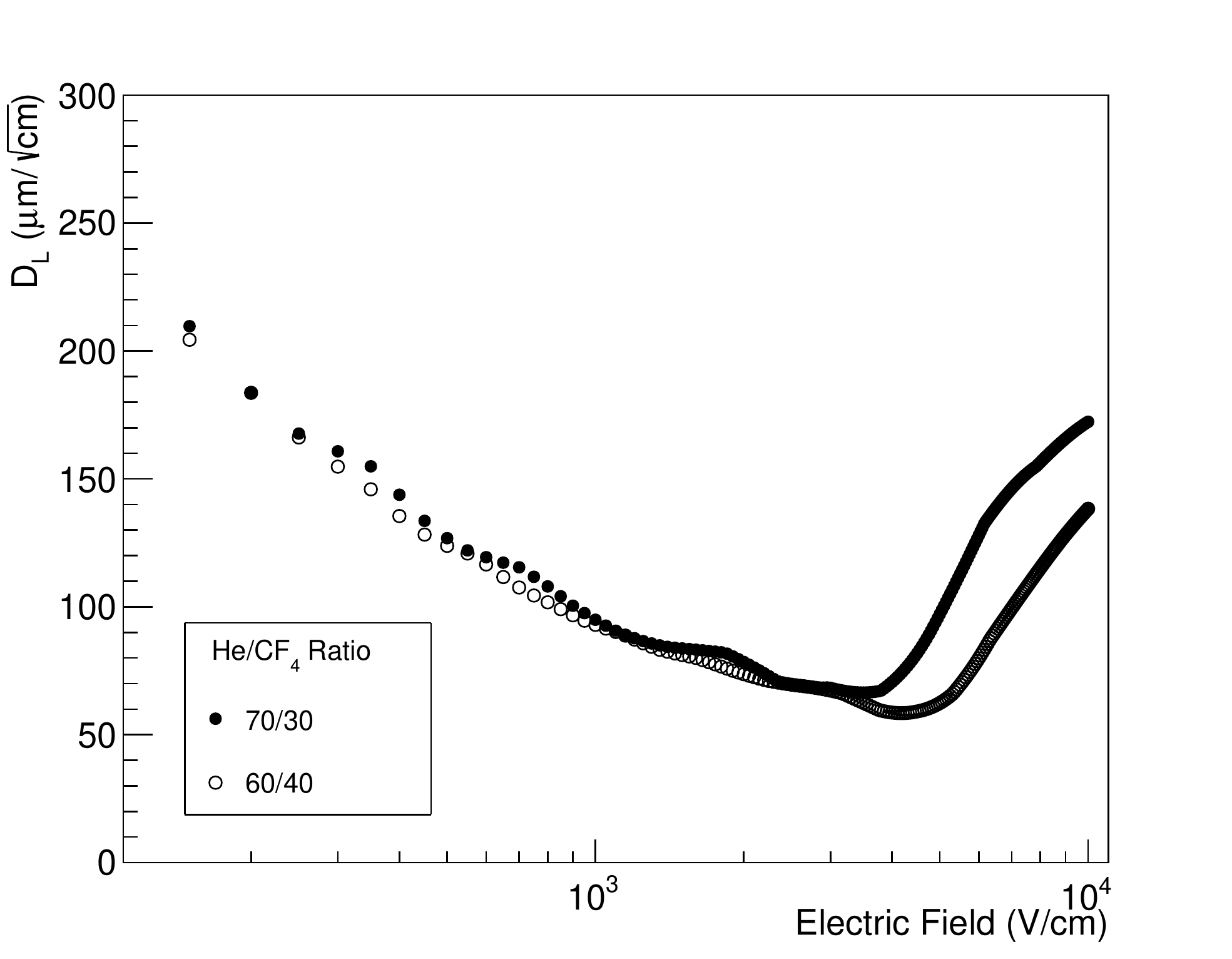}\\
\vspace{-0.3cm}
\mbox{a)} \hspace{5.5cm} \mbox{b)}
\vspace{-0.3cm}
\caption{Transverse (a) and longitudinal (b) diffusion coefficients for the two mixtures as a function of the electric field.}
\label{fig:diff}
\end{figure}

The behavior of the diffusion coefficients for different electric fields is reported in Fig.~\ref{fig:diff}
In particular, for an electric field of 0.5~kV/cm (i.e. the value set in the sensitive volume for the measurements presented in this paper) they were evaluated to be:
$$
D_{\mathrm{T}}^{60/40}~=~140~\frac{\mu{\mathrm{m}}}{\sqrt{\mathrm{cm}}}
{\mathrm{~~and~~}} 
D_{\mathrm{T}}^{70/30}~=~145~\frac{\mu{\mathrm{m}}}{\sqrt{\mathrm{cm}}}
$$

$$
D_{\mathrm{L}}^{60/40}~=~120~\frac{\mu{\mathrm{m}}}{\sqrt{\mathrm{cm}}}
{\mathrm{~~and~~}} 
D_{\mathrm{L}}^{70/30}~=~125~\frac{\mu{\mathrm{m}}}{\sqrt{\mathrm{cm}}}
$$

\begin{figure}[ht]
\centering
\includegraphics[width=0.45\textwidth]{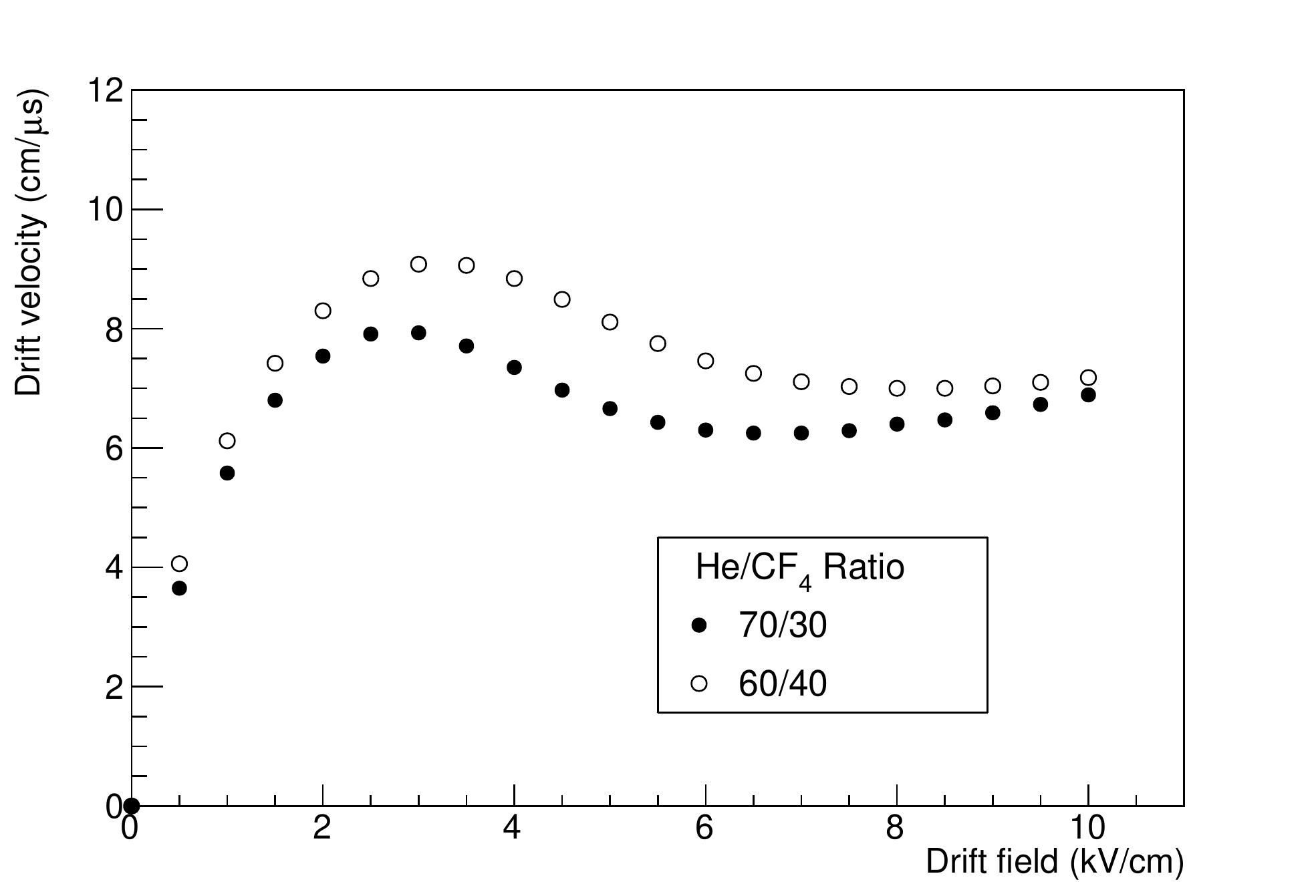}
\includegraphics[width=0.43\textwidth]{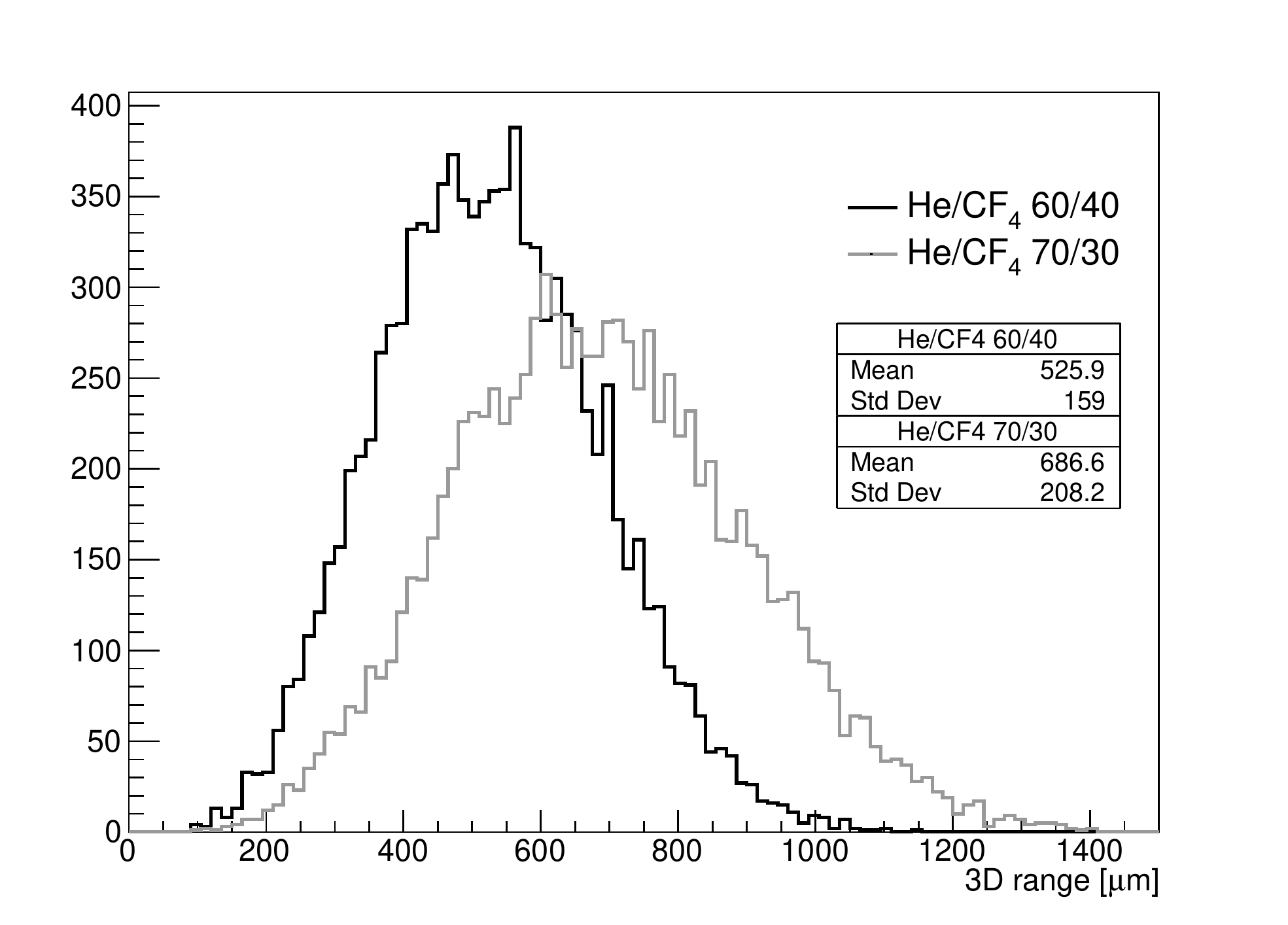}\\
\vspace{-0.3cm}
\mbox{a)} \hspace{6cm} \mbox{b)}
\vspace{-0.3cm}
\caption{Electron drift velocities as evaluated with Garfield for the two gas mixtures (a) and distribution of ranges of 5.9~keV electrons evaluated with GEANT4 (b) for the two gas mixtures}
\label{fig:drift_range}
\end{figure}

The electron drift velocity as a function of the electric field is shown on Fig.~\ref{fig:drift_range} (a).

It is clearly visible that a larger fraction of CF$_4$ allows to have slightly larger drift velocities. In particular, for an electric field of 500~V/cm:

$$
v^{60/40}_{\mathrm{drift}}~=~40.6~\mu{\mathrm{m/ns}}
{\mathrm{~~and~~}} 
v^{70/30}_{\mathrm{drift}}~=~36.5~\mu{\mathrm{m/ns}}
$$

Mean free path of the 5.9~keV electrons produced by photo-electric effect in gas by the 
$^{55}$Fe photons was evaluated with GEANT4~\cite{bib:geant} for the two gas mixtures.
The obtained distributions are shown in Fig.~\ref{fig:drift_range}~(b).

Even if a larger fraction of Helium increases the average path, in both cases, average ranges of several hundreds of micrometers were found. Therefore, interaction of $^{55}$Fe photons are expected to produce light spots with dimensions mainly due to the diffusion of electrons in the gas.

\section{Detector performance}

\begin{figure}[ht]
\begin{center}
\includegraphics[width=0.5\textwidth]{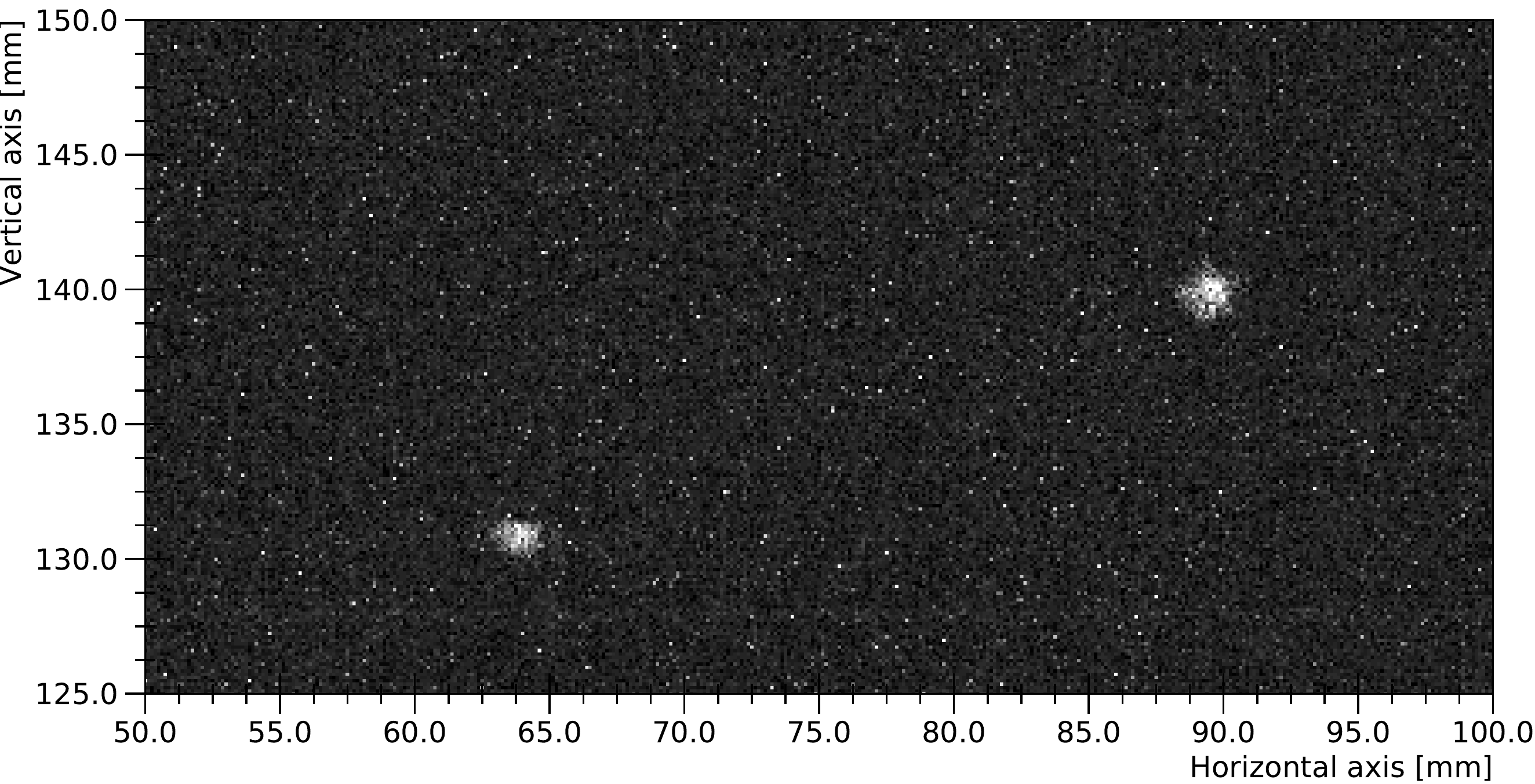}
\includegraphics[width=0.288\linewidth]{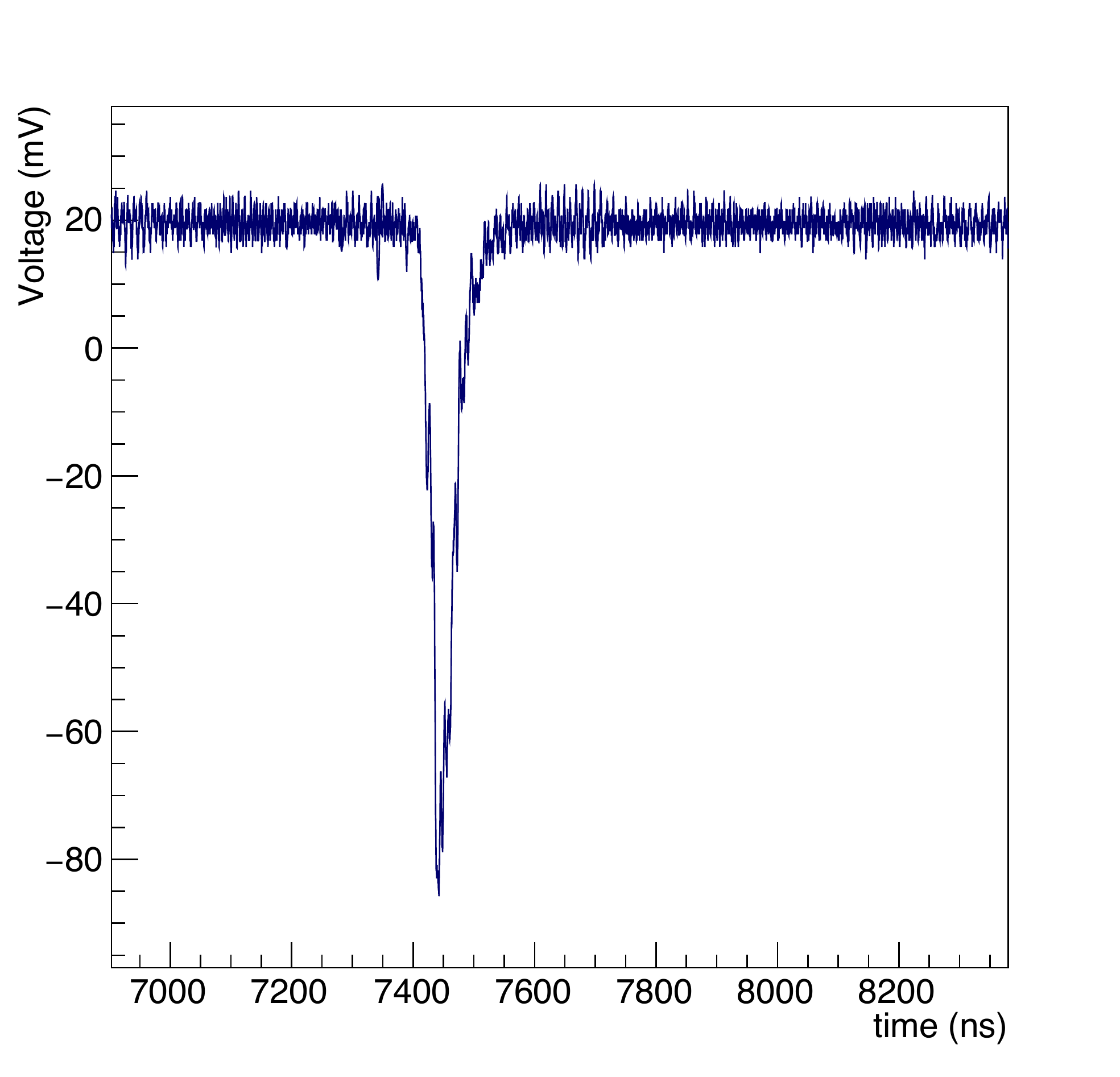}
\end{center}
\vspace{-0.6cm}
\hspace{5.5cm} \mbox{a)} \hspace{5.0cm} \mbox{b)}
\vspace{-0.3cm}

\caption{Detection of interactions of $^{55}$Fe photons in the sensitive volume. (a) sCMOS image with two light spots due to two photons interacting in the gas. (b) an example of PMT signal from a photon interaction.} 
\label{fig:spot}
\end{figure}

The detection performance provided by the two gas mixtures were evaluated by studying signals produced in the sensitive volume by 5.9~keV photons.
As described in \cite{bib:fe55}, the interactions of low energy photons with atoms in the gas mixtures create photo-electrons that release their whole energy in few hundreds of micrometers.
The subsequent diffusion of ionized electrons, makes these events produce round-like spots on the CMOS sensor with diameters of 2-3~mm as the ones shown in Fig.~\ref{fig:spot} due to two photons independently interacting in the gas.

\subsection{Data analysis}

Data acquired with CMOS camera were analysed by using the DBSCAN-based algorithm described in details in \cite{bib:algo}. In each image, clusters of illuminated pixels are individuated and used to reconstruct the signal spots.
For subsequent analysis, the position, the size and the total light of each spot are recorded.

For all events, a very simple analysis was performed on waveform provided by the PMT:
\begin{itemize}
    \item signal is integrated to evaluate the charge provided; 
    \item the FWHM of peak shape is measured to evaluate the signal duration.
\end{itemize}

\begin{figure}[ht]
\centering
\includegraphics[width=0.4\textwidth]{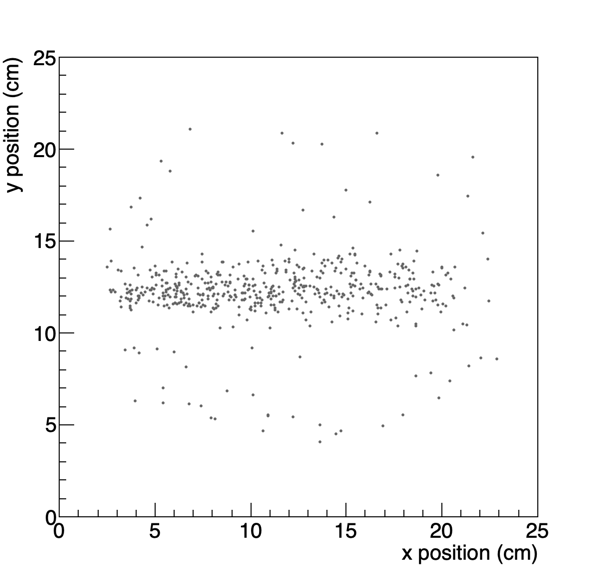}
\includegraphics[width=0.4\textwidth]{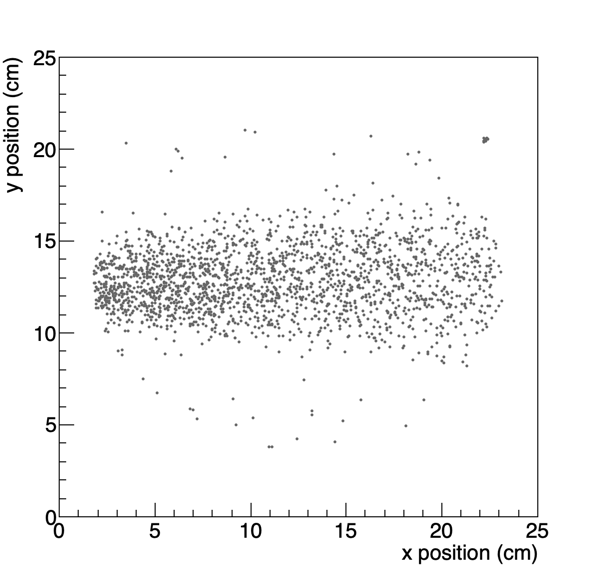}\\
\vspace{-0.3cm}
\mbox{a)} \hspace{6cm} \mbox{b)}
\vspace{-0.3cm}
\caption{Maps of the position of reconstructed $^{55}$Fe spots for 60/40 (a) and for 70/30 (b) in two typical runs.} 
\label{fig:map}
\end{figure}

The source was placed at a distance of 14~cm from border of the sensitive volume in a metal collimator that made it possible to limit the "illuminated" region to a cone. 
For most of following measurements the source was kept at a distance of 10.5~cm from the GEM stack. 
The maps of the positions of all reconstructed clusters in a typical run are shown in Fig.~\ref{fig:map}. 

The shape of the collimator hole was changed during the data taking. 
As a consequence, as it is well visible from the maps, the number of events due to interactions of 5.9~keV photons in the runs with the 70/30 (right panel in Fig.~\ref{fig:map}) is found to be almost five times larger with respect to the ones 60/40 (left panel in Fig.~\ref{fig:map}).

Due to the source position, 
a large number of clusters due to $^{55}$Fe photon interactions
is reconstructed in the central region of the sensitive volume 
(at an height between 10~cm and 15~cm).

\subsection{Light yield and energy resolution}

\subsubsection{Measurements with sCMOS}

\begin{figure}[ht]
\centering
\includegraphics[width=0.45\textwidth]{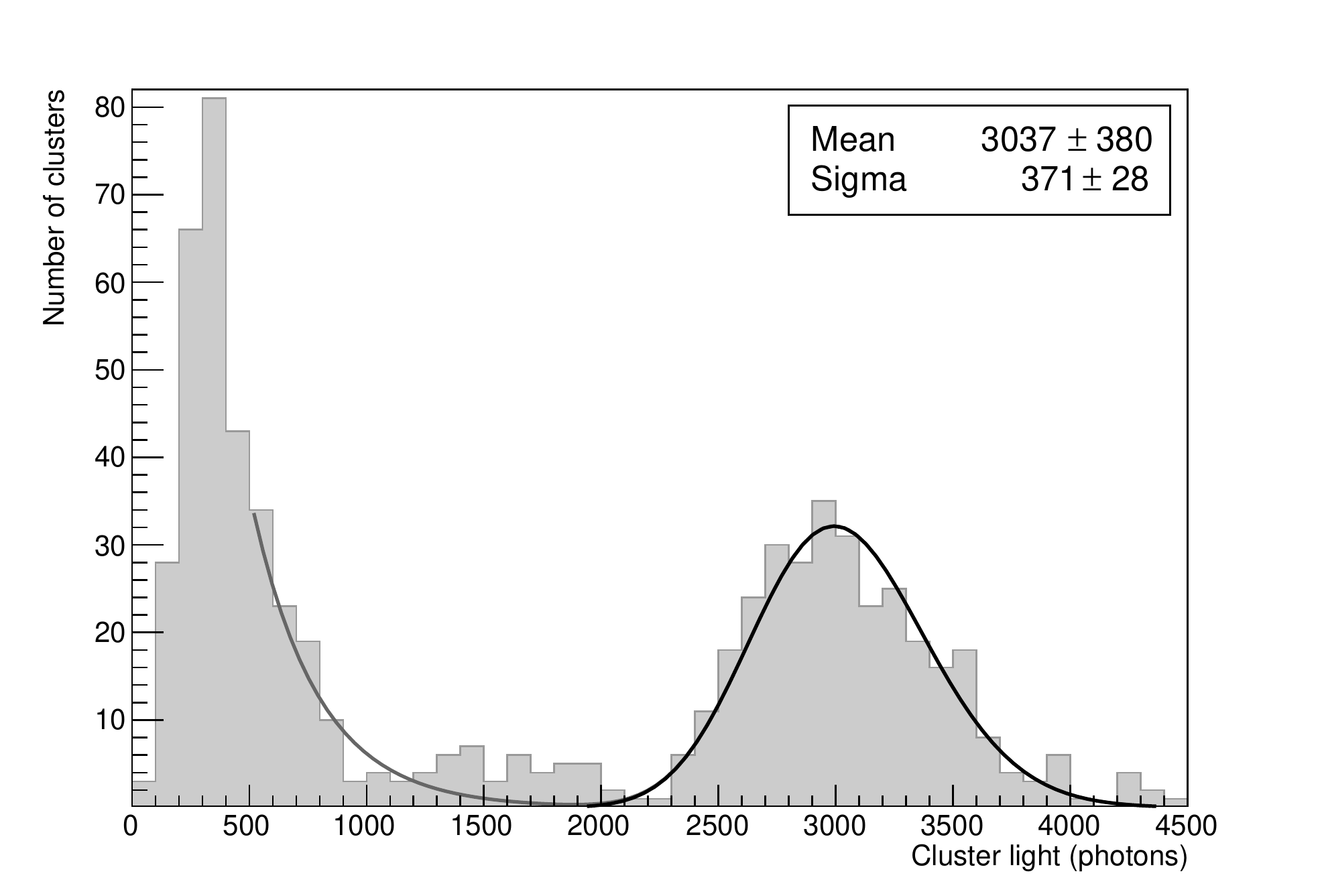}
\includegraphics[width=0.45\textwidth]{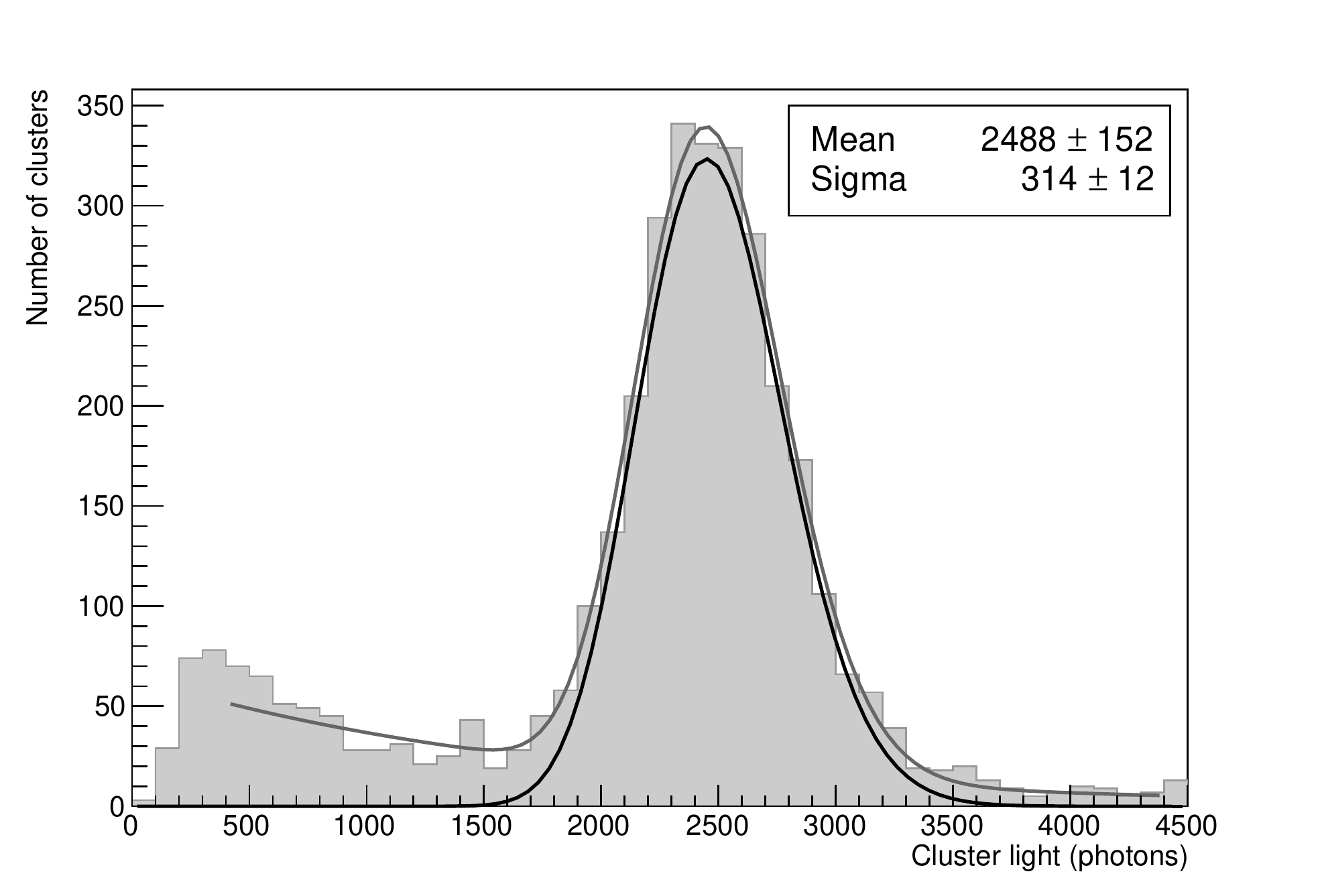}\\
\vspace{-0.3cm}
\mbox{a)} \hspace{6.5cm} \mbox{b)}
\vspace{-0.3cm}

\caption{Distribution of the light content in spots reconstructed for the 60/40 (a) and the 70/30 (b) gas mixtures with superimposed fit to an exponential plus a Polya function. The Polya contribution is also indicated.} 

\label{fig:peaks}
\end{figure}

Figure~\ref{fig:peaks} shows the spectra of the amount of light detected in spots reconstructed on sCMOS sensor.
The distributions are fitted to a function obtained by the sum of an exponential decay (to describe the background behavior) 
and a Polya~\cite{bib:rolandiblum} for the 5.9~keV peak.

As described in Sect~\ref{sec:oper}, in the chosen working conditions slightly different electron gains are expected, with a value 6.5\% larger for the 60/40.
Average light yields for the two mixtures were evaluated by dividing the average value of the Polya fit to the two distributions by 5.9~keV:
\begin{itemize}
    \item {\bf 60/40} provides an average value of 514~$\pm$~63 detected photons per keV released in the gas (in agreement with results obtained with lower \Vg\ and \Et\ \cite{bib:fe55}) with a relative fluctuation of 12.2\%;
    \item {\bf 70/30} provides an average value of 420~$\pm~$53 detected photons per keV released in the gas with a relative fluctuation of 12.6\%;
\end{itemize}

A light production 18\% larger for 60/40 was measured with respect to 70/30. The slightly lower light detected by the sCMOS with less CF$_4$ is in good agreement with expectations (see Sect.~\ref{sec:daq}) and confirms the component around 600~nm to be due to CF$^*_3$ dis-excitation.

\subsubsection{Measurements with PMT}

\begin{figure}[ht]
\centering
\includegraphics[width=0.45\textwidth]{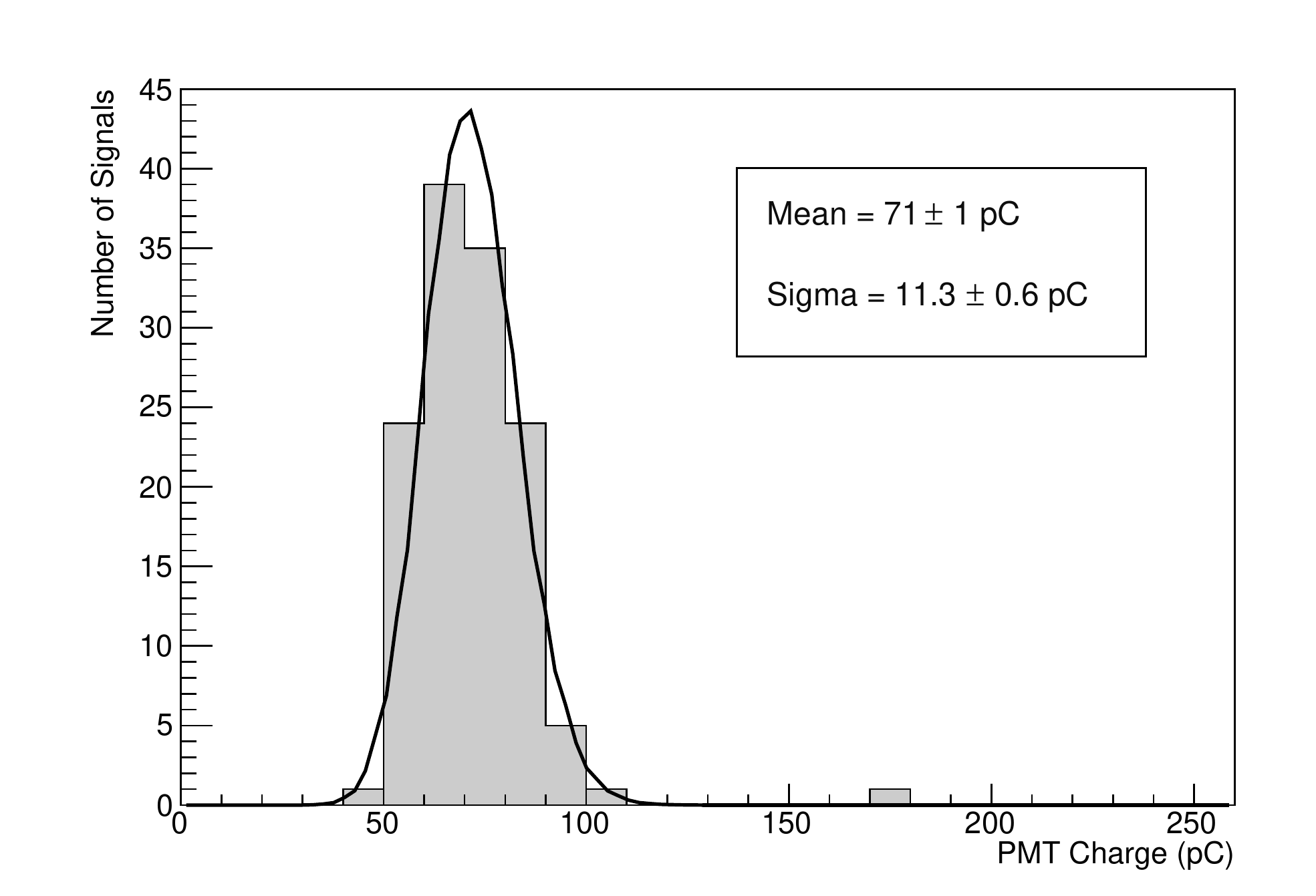}
\includegraphics[width=0.45\textwidth]{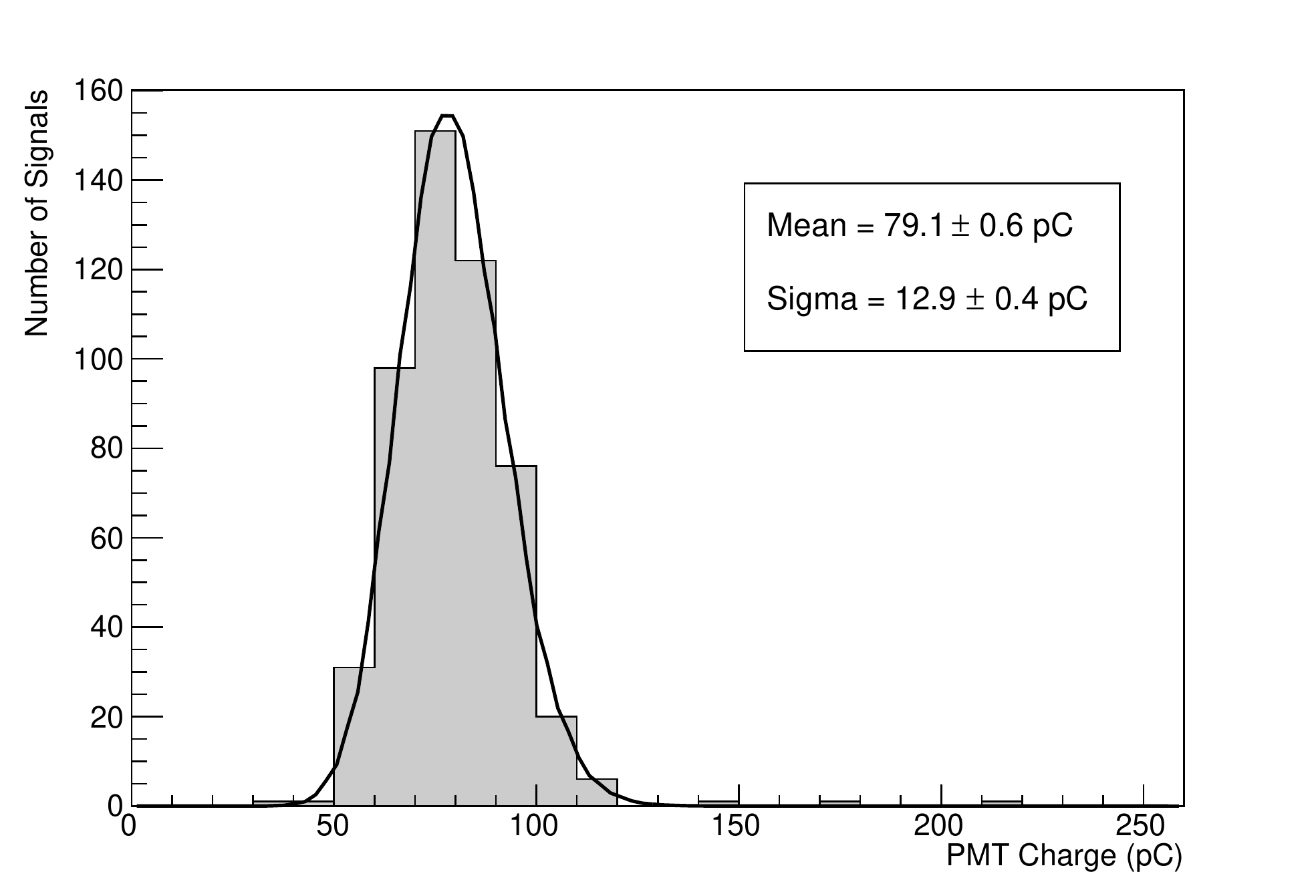}\\
\vspace{-0.3cm}
\mbox{a)} \hspace{5.5cm} \mbox{b)}
\vspace{-0.3cm}

\caption{Distribution of the charge provided by the PMT for the 60/40 (a) and the 70/30 (b) gas mixtures.} 
\label{fig:peaksPMT}
\end{figure}

The spectra of the charge integrated in Photo-multiplier waveforms
are shown in Figure~\ref{fig:peaksPMT} with a superimposed Polya fit.

The two distributions were fitted to a Polya function to evaluate the average light detected by the PMT with the two mixtures and 
light yield are obtained by dividing for 5.9~keV:
\begin{itemize}
    \item {\bf 60/40} provides an average value of (12.0~$\pm$~0.2) pC per keV released in the gas with a relative fluctuation of 15.5\%;
    \item {\bf 70/30} provides an average value of (13.4~$\pm$~0.1) pC per keV released in the gas with  a relative fluctuation of 16.3\%;
\end{itemize}

The PMT collects 12\% more light for 70/30 than 60/40. This can be explained by the larger amount of UV light produced by a more He rich mixture to whom PMT photo-cathode is sensitive (Sect.~\ref{sec:daq}).

\subsubsection{Energy resolution}

Since the energy release in gas is practically constant, the fluctuations of the response can be used to evaluate the resolution on the energy measurements at 5.9~keV. The results above, obtained with sCMOS and PMT indicate that with the sCMOS sensor a resolution of about 12\% is achieved with both mixtures, slightly better than what obtained with the PMT (between 15\% and 16\%) probably because of the very low noise level of the sCMOS sensor.
With both methods, a similar energy resolution was measured for the two gas mixtures confirming that the main contribution to this parameter is due to the statistical fluctuations of gas ionization and electron multiplication processes.

\subsubsection{Measurement of diffusion effect}

\begin{figure}[ht]
\centering
\includegraphics[width=0.45\textwidth]{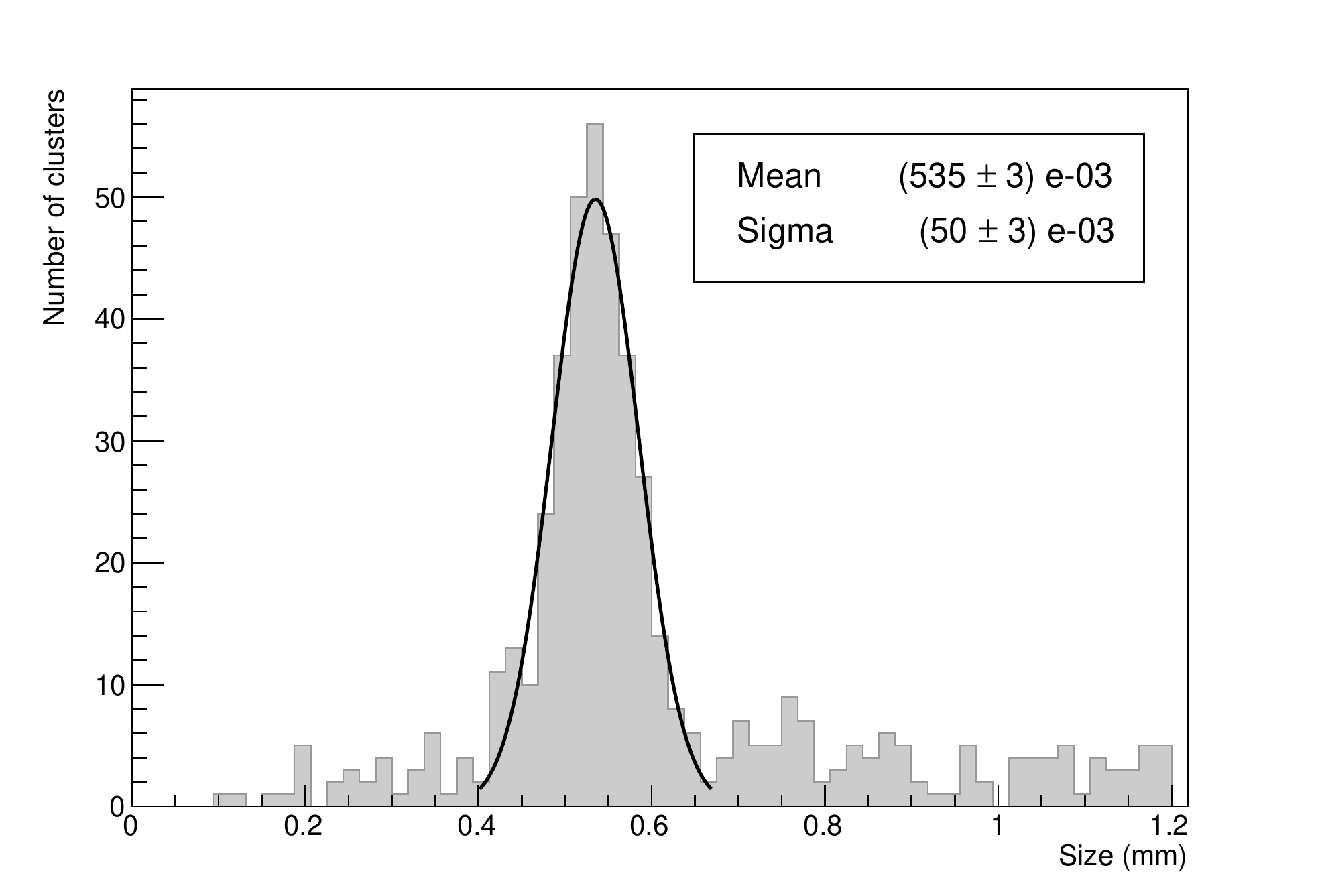}
\includegraphics[width=0.45\textwidth]{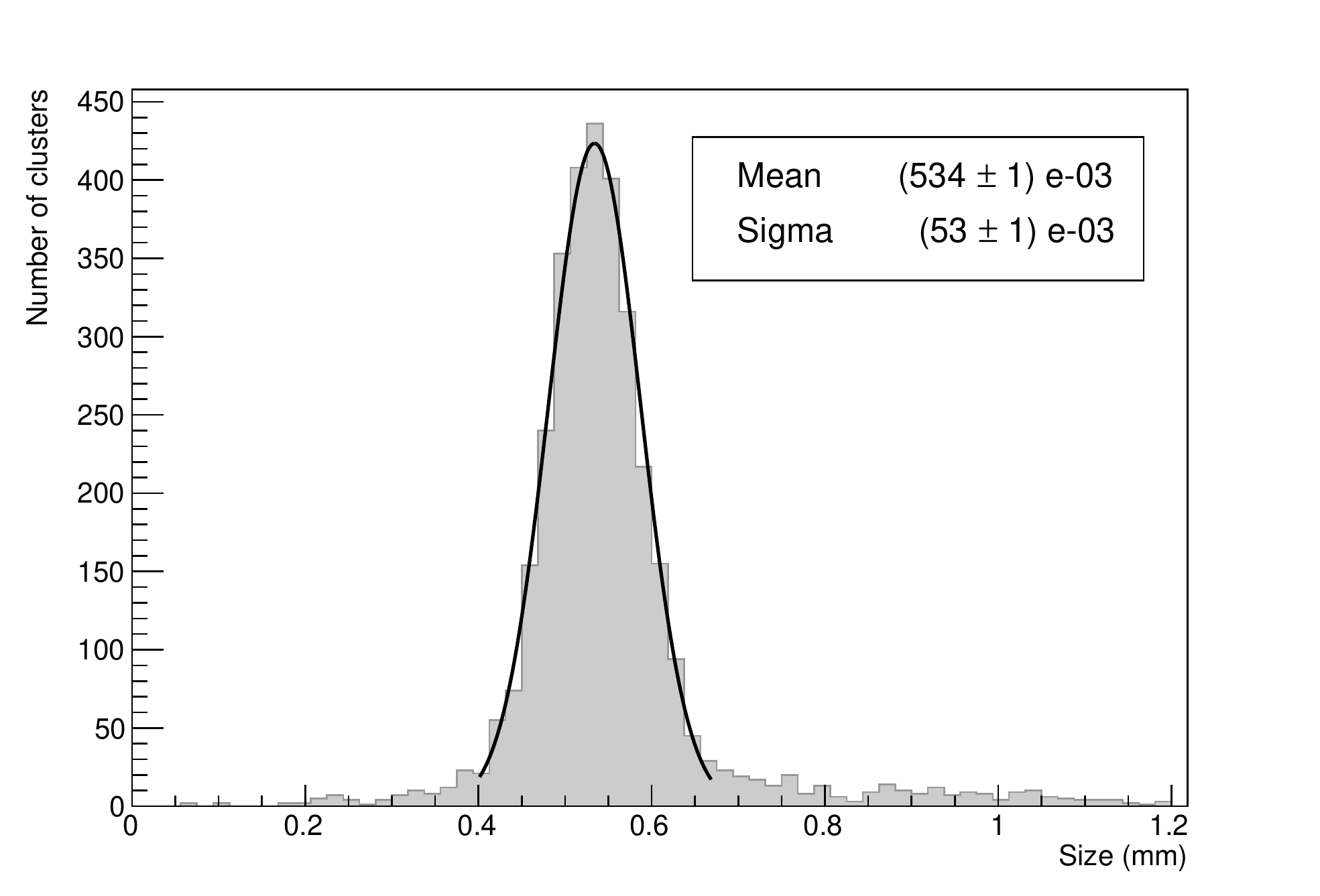} \\
\vspace{-0.3cm}
\mbox{a)} \hspace{6cm} \mbox{b)}
\vspace{-0.3cm}

\caption{Spectra of the value of $\sigma_T$ (see text for details) of spots reconstructed in the 60/40 (a) and the 70/30 (b) gas mixture.}
\label{fig:sizeZ}
\end{figure}

The distributions of $\sigma_{\mathrm{T}}$ for spots reconstructed with the source at a distance $z$ of 10.5~cm from the GEM were studied to evaluate the effect of the diffusion and are reported in Fig.~\ref{fig:sizeZ} with a superimposed gaussian fit.

The values of $\sigma_{\mathrm{T0}}$ were evaluated from Eq.~\ref{eq:diff} by using the diffusion coefficient given from the simulation (see Sect.~\ref{sect:simu}).
For the two mixtures they were found to be:
$$
\sigma^{60/40}_{\mathrm{T0}}~=~(280~\pm~60)~\mu{\mathrm{m}}
{\mathrm{~~and~~}} 
\sigma^{70/30}_{\mathrm{T0}}~=~(260~\pm~60)~\mu{\mathrm{m}}
$$
 
They are comparable within the measurement uncertainties confirming that they are mainly due to diffusion in the GEM stack.



%


\subsection{Detection efficiency}
\label{cap:detect}

The number of spots reconstructed as a function of the source position along $z$ axis was studied to evaluate the behavior of detection efficiency for interactions at different distances from the GEM.

\begin{figure}[ht]
\centering
\includegraphics[width=0.445\textwidth]{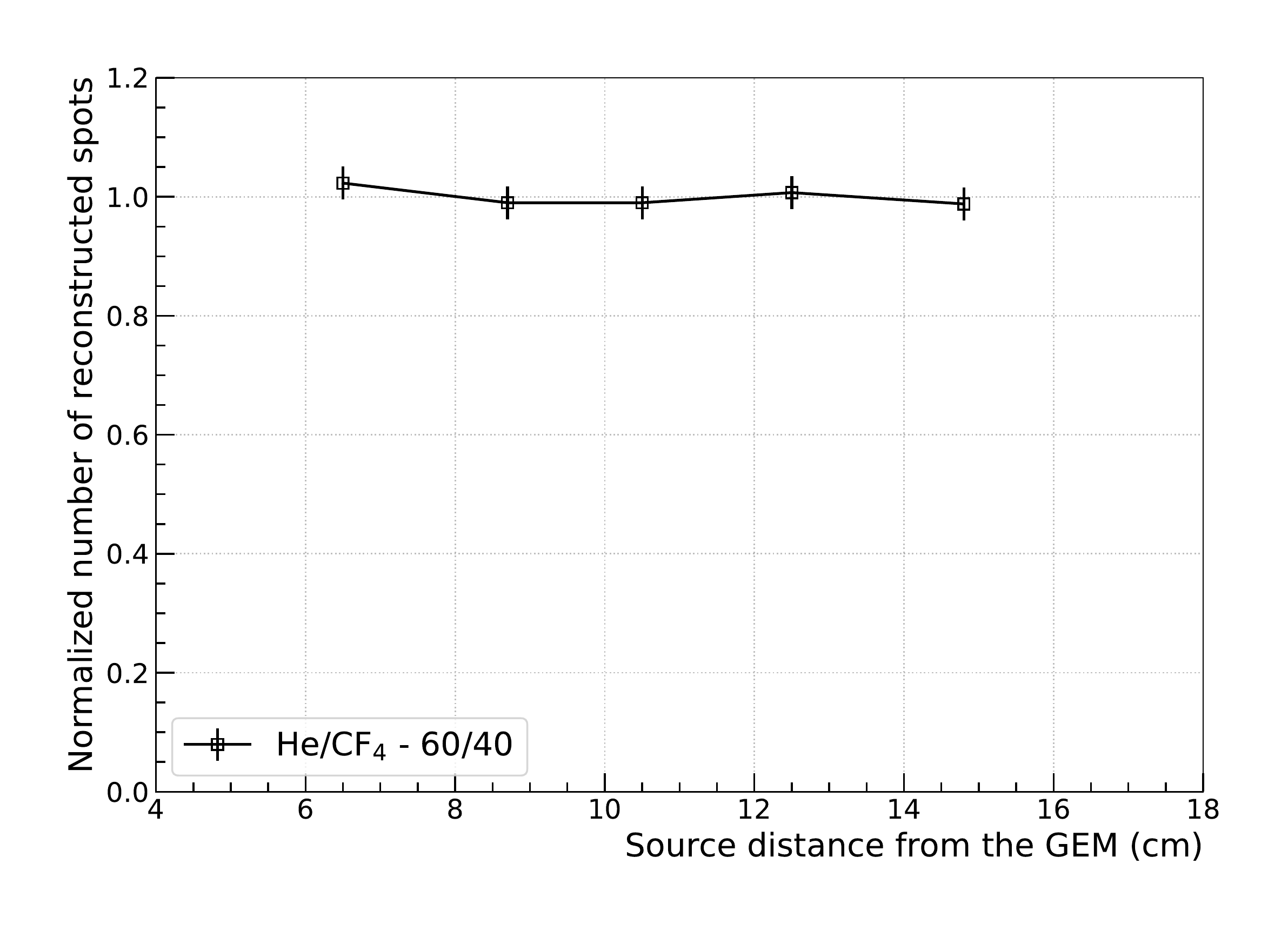}
\includegraphics[width=0.45\textwidth]{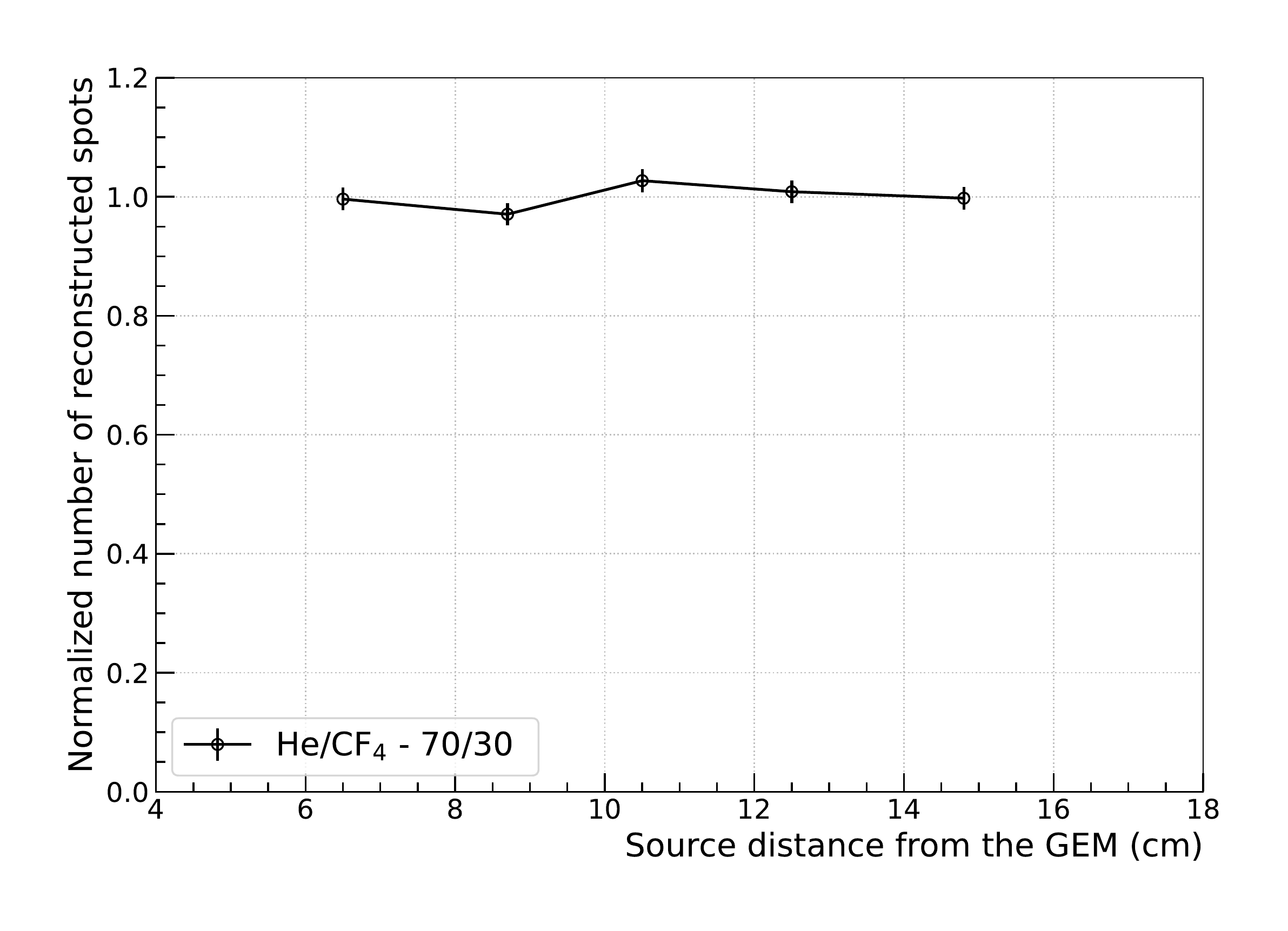}\\
\vspace{-0.3cm}
\mbox{a)} \hspace{6cm} \mbox{b)}
\vspace{-0.3cm}

\caption{Number of spots reconstructed (normalized to their average value) as a function of the position of the $^{55}$Fe source along the $z$ axis (a) for the 60/40 and (b) and 70/30).} 
\label{fig:effZ}
\end{figure}

Figure~\ref{fig:effZ} shows the behavior of the ratio between the number of reconstructed spots in the runs at different $z$ and and the average value in the whole scan.  
No evidence of a dependence of the detection efficiency on $z$ was found, allowing to conclude that, a constant detection efficiency is provided at all studied depths. This result is good in agreement with the very low electron absorption probability due to electron attachment estimated for He/CF$_4$  mixtures with Garfield for electric field values of \Ed\ used (see Sect.~\ref{sec:oper}).

\section{Conclusion}

The performance of an optical readout TPC with a sensitive volume of 7 litres was studied with two He/CF$_4$ based mixtures in different proportions (60/40 and 70/30).
The chosen detector electrical configurations, allowed to operate in very similar electron gas gain conditions.
Performed studies indicate that a constant detection efficiency was found in the sensitive volume 
together with a very good energy resolution around 13\% in both cases,
(even if the light yield for the 70/30 mixture resulted to be 18\% lesser) indicating that this is mainly due to ionization and multiplication statistics.
A sub-keV resolution is also very promising for application in Dark Matter search being the maximum energy released to an He nucleus by 1 GeV mass particle about 1 keV. 
Detector operation was monitored for a 25 days period. A detailed study of the behavior of currents and voltages provided by the supply system has shown the presence of two different kind instability events in GEM channels: discharges (with a sudden and fast current increase) and hot-spots (self sustaining events involving less current and creating small light spots on GEM surface).
Thanks to its quenching properties a larger presence of CF$_4$, showed to ensure a quite better stability with lower rate for both kinds of events.
Their occurrence was anyway found to be random in time with frequencies always lesser than few per hour and no evidence were found of correlation between two subsequent events.

Results presented in this paper demonstrated the possibility of operating the CYGNO prototype safe and stable conditions while providing promising performance (light yield, energy resolution and detection efficiency) in view of a larger TPC for dark matter search. 

\acknowledgments 
This work was supported by the European Research Council (ERC) under the European Union’s Horizon 2020 research and innovation program (grant agreement No 818744)”.

\bibliographystyle{JHEP}
\bibliography{LEMON-20-005}

\end{document}